\pdfoutput=1
\documentclass[12pt,a4paper]{article}
\usepackage{ifthen} % for conditional statements
\newboolean{pdflatex}
\setboolean{pdflatex}{true} % False for eps figures 

\newboolean{articletitles}
\setboolean{articletitles}{true} % False removes titles in references

\newboolean{uprightparticles}
\setboolean{uprightparticles}{false} %True for upright particle symbols

\def\paperauthors{LHCb collaboration} % Leave as is for PAPER, CONF and FIGURE
\def\paperasciititle{Unveiling the shape of the Neon-20 nucleus by measuring the flow coefficients with cumulants in PbNe and PbAr collisions at sqsnn = 70.9 GeV.} % Set ASCII title here !! MAKE sure it's only ASCII characters !! 
\def\papertitle{Unveiling the shape of the $^{20}$Ne nucleus by measuring the flow coefficients with cumulants in PbNe and PbAr collisions at $\sqsnn = 70.9$\gev} % Latex formatted title
\def\paperkeywords{{High Energy Physics}, {LHCb}} % Comma separated list
\def\papercopyright{\the\year\ CERN for the benefit of the LHCb collaboration} % new since 9/Apr/2018
\def\paperlicence{CC BY 4.0 licence}
\def\paperlicenceurl{https://creativecommons.org/licenses/by/4.0/}

\newif\ifEnableSectionTOCLinks
\EnableSectionTOCLinksfalse % deactivated

\usepackage[top=1in, bottom=1.25in, left=1in, right=1in]{geometry}

\usepackage{microtype}
\usepackage{lineno}  % for line numbering during review
\usepackage{xspace} % To avoid problems with missing or double spaces after
\usepackage{caption} %these three command get the figure and table captions automatically small

\usepackage{graphicx}  % to include figures (can also use other packages)
\usepackage{color}
\usepackage{colortbl}
\graphicspath{{./figs/}} % Make Latex search fig subdir for figures
\usepackage{amssymb}
\usepackage{amsfonts}
\usepackage{upgreek} % Adds in support for greek letters in roman typeset

\newcommand*\patchAmsMathEnvironmentForLineno[1]{%
\expandafter\let\csname old#1\expandafter\endcsname\csname #1\endcsname
\expandafter\let\csname oldend#1\expandafter\endcsname\csname
end#1\endcsname
 \renewenvironment{#1}%
   {\linenomath\csname old#1\endcsname}%
   {\csname oldend#1\endcsname\endlinenomath}%
}
\newcommand*\patchBothAmsMathEnvironmentsForLineno[1]{%
  \patchAmsMathEnvironmentForLineno{#1}%
  \patchAmsMathEnvironmentForLineno{#1*}%
}
\AtBeginDocument{%
\patchBothAmsMathEnvironmentsForLineno{equation}%
\patchBothAmsMathEnvironmentsForLineno{align}%
\patchBothAmsMathEnvironmentsForLineno{flalign}%
\patchBothAmsMathEnvironmentsForLineno{alignat}%
\patchBothAmsMathEnvironmentsForLineno{gather}%
\patchBothAmsMathEnvironmentsForLineno{multline}%
\patchBothAmsMathEnvironmentsForLineno{eqnarray}%
}

\usepackage[pdftex,
            pdfauthor={\paperauthors},
            pdftitle={\paperasciititle},
            pdfkeywords={\paperkeywords}]{hyperref}
\usepackage{hyperxmp}
\hypersetup{
    pdfcopyright={Copyright (C) \papercopyright},
    pdflicenseurl={\paperlicenceurl}
}
\usepackage[colorinlistoftodos,textsize=scriptsize]{todonotes}

\usepackage[bottom,flushmargin,hang,multiple]{footmisc}

\usepackage[all]{hypcap} % Internal hyperlinks to floats.

\usepackage{xspace} 
\usepackage{upgreek}

\def\lhcb   {\mbox{LHCb}\xspace}

\def\MagUp {\mbox{\em Mag\kern -0.05em Up}\xspace}

\ifthenelse{\boolean{uprightparticles}}%
{

 \def\PDelta      {\ensuremath{\Delta}\xspace}                 
 \def\PXi         {\ensuremath{\Xi}\xspace}                 
 \def\PLambda     {\ensuremath{\Lambda}\xspace}                 
 \def\PSigma      {\ensuremath{\Sigma}\xspace}                 
 \def\POmega      {\ensuremath{\Omega}\xspace}                 
 \def\PUpsilon    {\ensuremath{\Upsilon}\xspace}
 \let\oldPi\Pi
 \def\PPi         {\ensuremath{\oldPi}\xspace}

 \def\PB      {\ensuremath{\mathrm{B}}\xspace}                 
 \def\PD      {\ensuremath{\mathrm{D}}\xspace}                 

 \def\PK      {\ensuremath{\mathrm{K}}\xspace}                 
 \def\Ps      {\ensuremath{\mathrm{s}}\xspace}

 \def\thebaroffset{0.0em}
}
{

 \mathchardef\PDelta="7101
 \mathchardef\PXi="7104
 \mathchardef\PLambda="7103
 \mathchardef\PSigma="7106
 \mathchardef\POmega="710A
 \mathchardef\PUpsilon="7107
 \mathchardef\PPi="7105
 \def\PB      {\ensuremath{B}\xspace}                 
 \def\PD      {\ensuremath{D}\xspace}                 

 \def\PK      {\ensuremath{K}\xspace}                 
 \def\Ps      {\ensuremath{s}\xspace}

 \def\thebaroffset{0.18em}
}
\newcommand{\offsetoverline}[2][\thebaroffset]{\kern #1\overline{\kern -#1 #2}}%

\makeatletter
\ifcase \@ptsize \relax% 10pt
  \newcommand{\miniscule}{\@setfontsize\miniscule{4}{5}}% \tiny: 5/6
\or% 11pt
  \newcommand{\miniscule}{\@setfontsize\miniscule{5}{6}}% \tiny: 6/7
\or% 12pt
  \newcommand{\miniscule}{\@setfontsize\miniscule{5}{6}}% \tiny: 6/7
\fi
\makeatother

\DeclareRobustCommand{\optbar}[1]{\shortstack{{\miniscule (\rule[.5ex]{1.25em}{.18mm})}
  \\ [-.7ex] $#1$}}

\def\squark    {{\ensuremath{\Ps}}\xspace}

\def\KorKbar {\kern \thebaroffset\optbar{\kern -\thebaroffset \PK}{}\xspace}

\def\D       {{\ensuremath{\PD}}\xspace}

\def\DorDbar {\kern \thebaroffset\optbar{\kern -\thebaroffset \PD}\xspace}

\def\Dp      {{\ensuremath{\D^+}}\xspace}
\def\Dm      {{\ensuremath{\D^-}}\xspace}

\def\DpDm    {\ensuremath{\Dp {\kern -0.16em \Dm}}\xspace}

\def\B       {{\ensuremath{\PB}}\xspace}

\def\BorBbar {\kern \thebaroffset\optbar{\kern -\thebaroffset \PB}\xspace}

\def\Bd      {{\ensuremath{\B^0}}\xspace}

\def\BdorBdbar {\kern \thebaroffset\optbar{\kern -\thebaroffset \Bd}\xspace}

\def\Bs      {{\ensuremath{\B^0_\squark}}\xspace}

\def\BsorBsbar {\kern \thebaroffset\optbar{\kern -\thebaroffset \Bs}\xspace}

\def\Y#1S{\ensuremath{\PUpsilon{(#1S)}}\xspace}

\def\LorLbar     {\kern \thebaroffset\optbar{\kern -\thebaroffset \PLambda}\xspace}

\def\AT#1     {\ensuremath{A_{\mathrm{T}}^{#1}}\xspace}           % 2

\def\C#1      {\ensuremath{\mathcal{C}_{#1}}\xspace}                       % 9
\def\Cp#1     {\ensuremath{\mathcal{C}_{#1}^{'}}\xspace}                    % 7
\def\Ceff#1   {\ensuremath{\mathcal{C}_{#1}^{\mathrm{(eff)}}}\xspace}        % 9  
\def\Cpeff#1  {\ensuremath{\mathcal{C}_{#1}^{'\mathrm{(eff)}}}\xspace}       % 7
\def\Ope#1    {\ensuremath{\mathcal{O}_{#1}}\xspace}                       % 2
\def\Opep#1   {\ensuremath{\mathcal{O}_{#1}^{'}}\xspace}                    % 7

\newcommand{\nospaceunit}[1]{\ensuremath{\text{#1}}}       
\newcommand{\aunit}[1]{\ensuremath{\text{\,#1}}}       
\newcommand{\tev}{\aunit{Te\kern -0.1em V}\xspace}
\newcommand{\gev}{\aunit{Ge\kern -0.1em V}\xspace}
\newcommand{\mev}{\aunit{Me\kern -0.1em V}\xspace}
\newcommand{\kev}{\aunit{ke\kern -0.1em V}\xspace}
\newcommand{\ev}{\aunit{e\kern -0.1em V}\xspace}
 
\newcommand{\mevc}{\ensuremath{\aunit{Me\kern -0.1em V\!/}c}\xspace}
\newcommand{\gevc}{\ensuremath{\aunit{Ge\kern -0.1em V\!/}c}\xspace}
\newcommand{\mevcc}{\ensuremath{\aunit{Me\kern -0.1em V\!/}c^2}\xspace}
\newcommand{\gevcc}{\ensuremath{\aunit{Ge\kern -0.1em V\!/}c^2}\xspace}
\def\cm   {\aunit{cm}\xspace}

\def\mm   {\aunit{mm}\xspace}

\def\mub{\ensuremath{\,\upmu\nospaceunit{b}}\xspace}
\def\nb {\aunit{nb}\xspace}
\def\invnb {\ensuremath{\nb^{-1}}\xspace}

\newcommand{\chisq}{\ensuremath{\chi^2}\xspace}

\def\gsim{{~\raise.15em\hbox{$>$}\kern-.85em
          \lower.35em\hbox{$\sim$}~}\xspace}
\def\lsim{{~\raise.15em\hbox{$<$}\kern-.85em
          \lower.35em\hbox{$\sim$}~}\xspace}

\def\sqsnn {\ensuremath{\protect\sqrt{s_{\scriptscriptstyle\text{NN}}}}\xspace}
\def\pt         {\ensuremath{p_{\mathrm{T}}}\xspace}

\def\mrad{\aunit{mrad}\xspace}

\def\evtgen     {\mbox{\textsc{EvtGen}}\xspace}

\def\geant      {\mbox{\textsc{Geant4}}\xspace}

\def\photos     {\mbox{\textsc{Photos}}\xspace}

\def\tell1  {TELL1\xspace}
\def\ukl1   {UKL1\xspace}

\newcommand{\ie}{\mbox{\itshape i.e.}\xspace}

\newcommand{\lhcborcid}[1]{\href{https://orcid.org/#1}{\hspace*{0.1em}\raisebox{-0.45ex}{\includegraphics[width=1em]{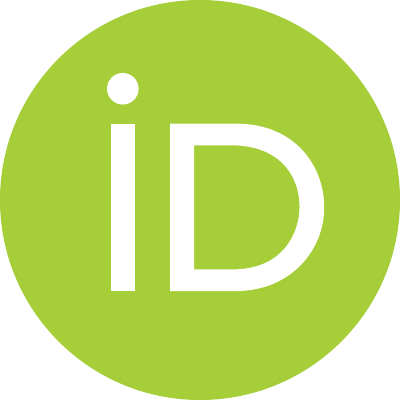}}}}

\hypersetup{
  colorlinks   = true, %Colours links instead of ugly boxes
  urlcolor     = blue, %Colour for external hyperlinks
  linkcolor    = blue, %Colour of internal links
  citecolor    = red   %Colour of citations
}

\ifEnableSectionTOCLinks
    \usepackage[explicit]{titlesec} % to change headings
    
    \let\oldcontentsline\contentsline
    \renewcommand

    \titleformat{\section}{\normalfont\Large\bf}{\hyperlink{tocsection.\thesection}{{\thesection} \parbox[t]{\dimexpr\textwidth-1pc}{#1}}}{1pc}{}

    \titleformat{\subsection}{\normalfont\bf}{\hyperlink{tocsubsection.\thesubsection}{{\thesubsection} \parbox[t]{\dimexpr\textwidth-1pc}{#1}}}{1pc}{}

    \titleformat{name=\section,numberless}[display]{}{}{0pt}{\normalfont\Huge\bfseries #1}
\fi

\usepackage{cite} % Allows for ranges in citations
\usepackage{LHCb/mciteplus}
\usepackage{longtable} % only for template; not usually to be used in PAPERs
\usepackage{booktabs}
\usepackage{multirow}
\usepackage{amsmath} % Adds a large collection of math symbols

\newcommand{\NVPClusters}{\ensuremath{N_{\text{VP}}^{\text{Clusters}}}\xspace}

\newcommand{\EcalE}{\ensuremath{E_{\text{tot}}^{\text{ECAL}}}\xspace}
\newcommand{\clonesangle}{\ensuremath{\theta_{\text{angle}}}\xspace}

\def\invub {\ensuremath{\mub^{-1}}\xspace}

\newcommand{\vtwo}{\ensuremath{v_2}\xspace}
\newcommand{\vthree}{\ensuremath{v_3}\xspace}
\newcommand{\vtwotwo}{\ensuremath{v_2\{2\}}\xspace}
\newcommand{\vthreetwo}{\ensuremath{v_3\{2\}}\xspace}
\newcommand{\vtwofour}{\ensuremath{v_2\{4\}}\xspace}

\newcommand{\cproxy}{\ensuremath{q_c(\EcalE)}\xspace}

\begin{document}

%%%%%%%%%%%%%%%%%%%%%%%%%
%%%%% Title     %%%%%%%%%
%%%%%%%%%%%%%%%%%%%%%%%%%
\renewcommand{\thefootnote}{\fnsymbol{footnote}}
\setcounter{footnote}{1}

% %%%%%%% CHOOSE TITLE PAGE--------
%\onecolumn
% ===============================================================================
% Purpose: LHCb-CONF Note title page template
% Author: 
% Created on: 2010-09-25
% ===============================================================================

%%%%%%%%%%%%%%%%%%%%%%%%%
%%%%%  TITLE PAGE  %%%%%%
%%%%%%%%%%%%%%%%%%%%%%%%%
\begin{titlepage}

% Header ---------------------------------------------------
\vspace*{-1.5cm}

\noindent
\begin{tabular*}{\linewidth}{lc@{\extracolsep{\fill}}r@{\extracolsep{0pt}}}
\ifthenelse{\boolean{pdflatex}}% Logo format choice
{\vspace*{-1.2cm}\mbox{\!\!\!\includegraphics[width=.14\textwidth]{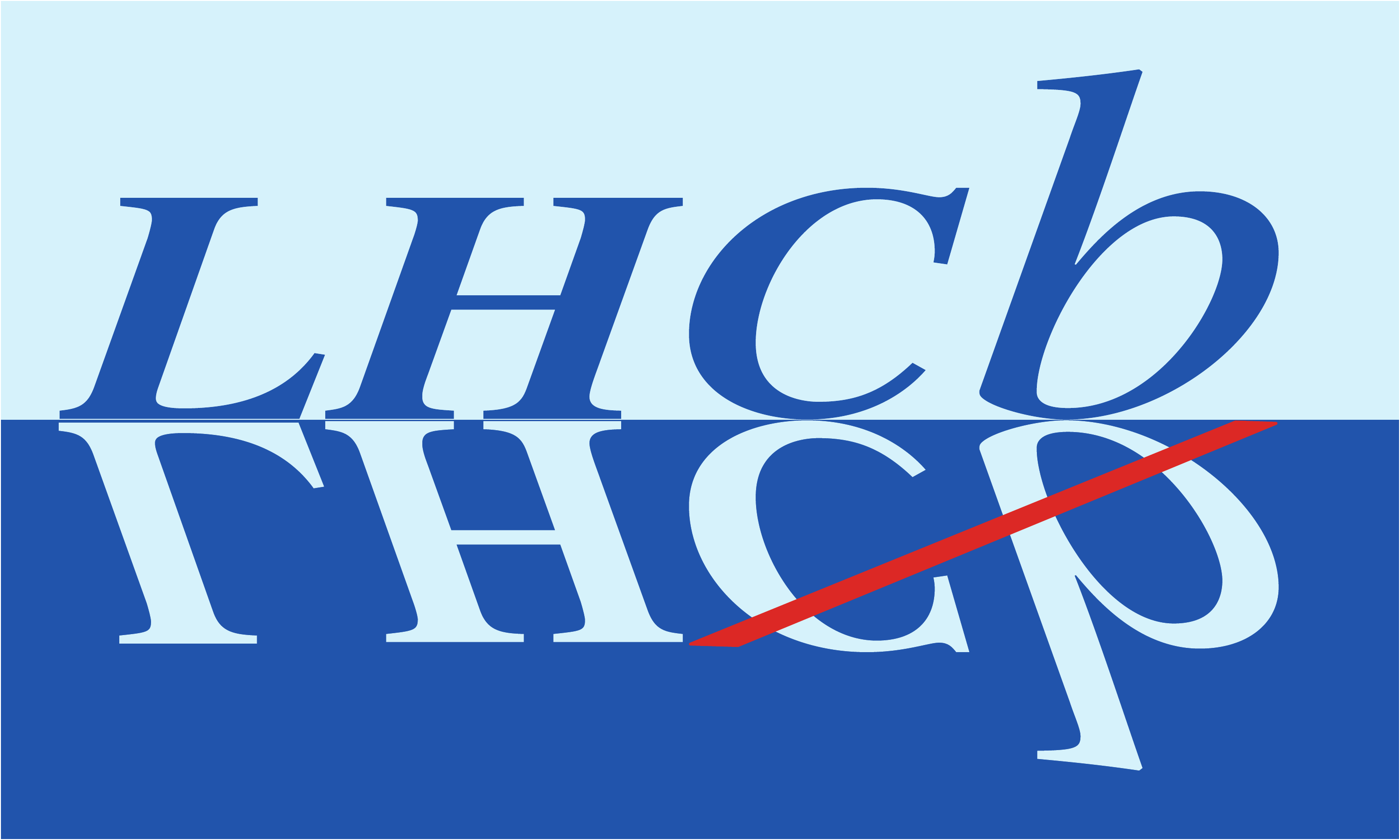}} & &}%
{\vspace*{-1.2cm}\mbox{\!\!\!\includegraphics[width=.12\textwidth]{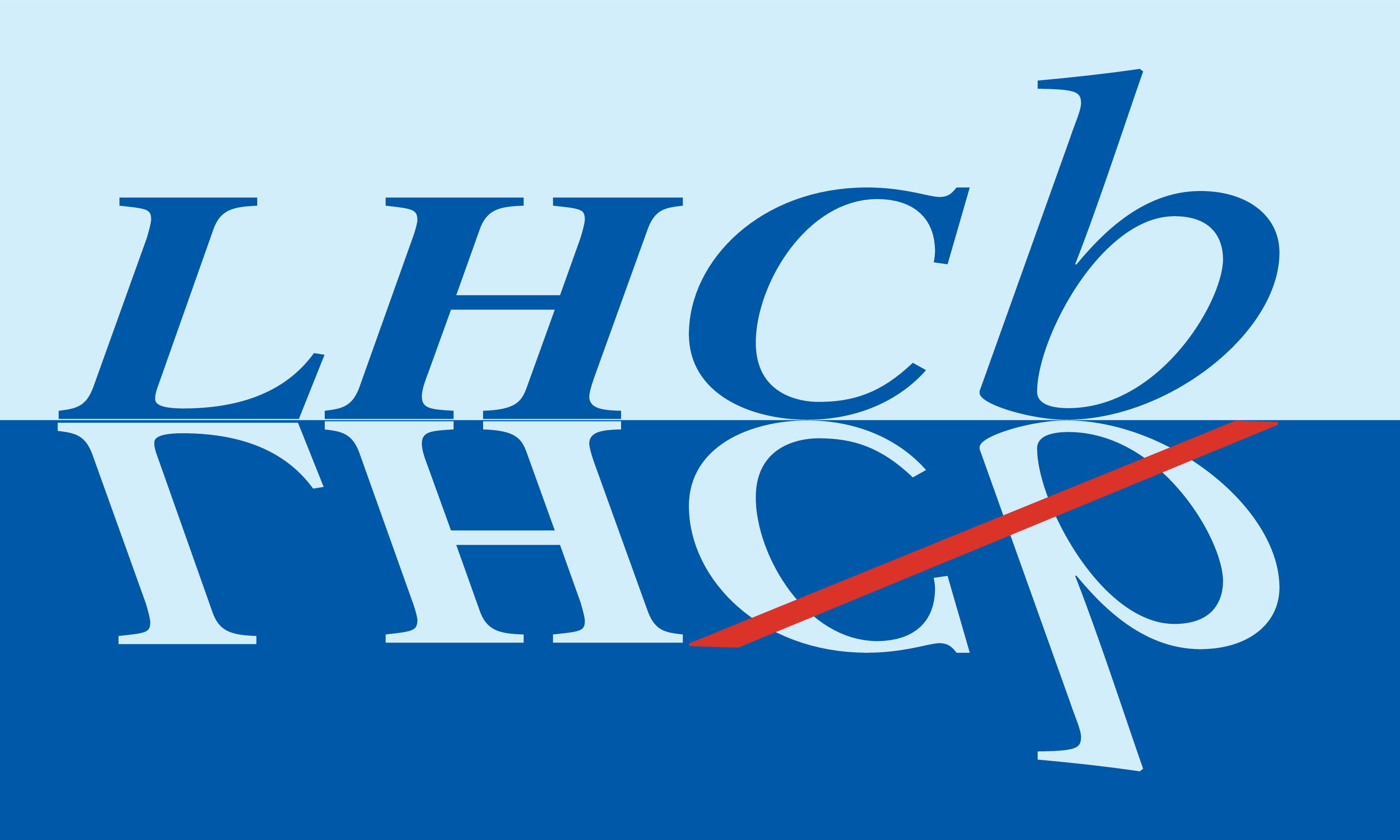}} & &}
 \\
 & & LHCb-CONF-2025-001 \\  % ID 
 & & September 15, 2025 \\ % Date - Can also hardwire e.g.: 23 March 2010
 & & \\
\hline
\end{tabular*}

\vspace*{1.8cm}

% Title --------------------------------------------------
{\normalfont\bfseries\boldmath\huge
\begin{center}
% DO NOT EDIT HERE. Instead edit macro in main.tex to keep metadata correct
  \papertitle
\end{center}
}

\vspace*{1.0cm}

% Authors -------------------------------------------------
\begin{center}
\paperauthors % Edit macro in main.tex to keep metadata correct
   % Identify conference in the footnote
   \footnote{Conference report prepared for the 8$^{\rm th}$ International Conference on the Initial Stages of High-Energy Nuclear Collisions, Taipei, 2025.
   % Edit to contain the names of the one or two proponents
   Contact author: Zhengchen Lian, 
   \href{mailto:zhengchen.lian@cern.ch}{zhengchen.lian@cern.ch}
 }
\end{center}

\vspace{\fill}

% Abstract -----------------------------------------------
\begin{abstract}
  \noindent  The anisotropic flow coefficients $v_n$ quantify the collective medium response to the initial spatial anisotropy of the overlapping region in ion collisions and serve as sensitive probes of both the medium properties and shape of nuclear initial states. In this analysis, the \vtwo and \vthree parameters of prompt charged particles are measured using the multiparticle cumulant method in fixed-target PbNe and PbAr collisions at $\sqsnn = 70.9\gev$, collected by LHCb using the SMOG2 gas-target system during the 2024 LHC lead-beam run. The cumulant method is first validated using 2018 PbPb collision data, successfully reproducing previous measurements obtained via the two-particle correlation method. Results for the fixed-target collisions are then presented, showing a significantly larger value of the elliptic flow coefficient \vtwo in central PbNe with respect to PbAr collisions. This is qualitatively consistent with 3+1D hydrodynamic predictions including \textit{ab-initio} descriptions of the nuclear structure. The results provide the first experimental confirmation of the distinctive bowling-pin shape of the ground-state $^{20}$Ne nucleus, validating at the same time the hydrodynamic description of the hot medium formed in high-energy collisions involving light ions. 
\end{abstract}

\vspace*{2.0cm}
\vspace{\fill}
{\footnotesize
% Edit macro in main.tex to keep metadata correct
\centerline{\copyright~\papercopyright. \href{\paperlicenceurl}{\paperlicence}.}}
\vspace*{2mm}

\end{titlepage}

\pagestyle{empty}  % no page number for the title 

%%%%%%%%%%%%%%%%%%%%%%%%%%%%%%%%
%%%%%  EOD OF TITLE PAGE  %%%%%%
%%%%%%%%%%%%%%%%%%%%%%%%%%%%%%%%

%  empty page follows the title page ----
\newpage
\setcounter{page}{2}
\mbox{~}

%\twocolumn
% %%%%%%%%%%%%% ---------

\renewcommand{\thefootnote}{\arabic{footnote}}
\setcounter{footnote}{0}

%%%%%%%%%%%%%%%%%%%%%%%%%%%%%%%%
%%%%%  Table of Content   %%%%%%
%%%%%%%%%%%%%%%%%%%%%%%%%%%%%%%%
%%%% Uncomment if desired
%\tableofcontents

\cleardoublepage

%%%%%%%%%%%%%%%%%%%%%%%%%
%%%%% Main text %%%%%%%%%
%%%%%%%%%%%%%%%%%%%%%%%%%

\pagestyle{plain} % restore page numbers for the main text
\setcounter{page}{1}
\pagenumbering{arabic}

%% Uncomment during review phase. 
%% Comment before a final submission.
% \linenumbers

%% This is the main body
%% It is useful to have a single file so comments are not missed in overleaf.
\section{Introduction}
\label{sec:Introduction}
According to hydrodynamic models of heavy-ion collisions at relativistic energy, the collective flow coefficients, extensively studied by experiments at RHIC~\cite{PHENIX_flow, STAR_flow} and LHC~\cite{ALICE:2019zfl, ATLAS_flow, CMS_flow, LHCb-PAPER-2023-031} accelerators, are highly sensitive to the initial conditions of the collision, such as the shape of the overlapping region and the fluctuations in the energy density~\cite{Shape_nuclei_theory}. 
The sensitivity of these observables to the nuclear deformation of the colliding ions has been recently demonstrated, notably with the observation by the STAR collaboration of a large deformation in collisions of ground-state $^{238}\mathrm{U}$ nuclei, known for their elongated, axial-symmetric shape~\cite{STAR:2024wgy}.
In collisions involving light ions, the study of collective flow offers a unique tool to explore 
their nuclear structure, that is hardly accessible using traditional spectroscopic techniques at low energies,
limited by long-timescale quantum fluctuations.  
At the same time, it provides quantitative tests
of the hydrodynamic paradigm for system sizes where the formation of a quark-gluon plasma (QGP) 
is still debated. A notable example was given by the study of \textit{p}Au, \textit{d}Au, and $^3$HeAu collisions at RHIC~\cite{aidala_creation_2019,STAR:2022pfn}. 

The ground state of the $^{20}\mathrm{Ne}$ nucleus is particularly interesting because of its expected  
extreme deformation in a reflection-asymmetric $\alpha+^{16}$O or 5-$\alpha$-cluster molecular configuration\cite{Horiuchi:1968,Ebran:2012ww,Zhou:2013ala,ZHOU2016227,Freer:2017gip}. 
Combined with a small number of nucleons and relatively
small nuclear effects, this deformation is so large that its effects
can survive the QGP evolution and be observed in the final state. A
possible experimental strategy to test the initial-state modelling and hydrodynamic response in small systems consists of comparing
observables obtained from $^{20}\mathrm{Ne}^{20}\mathrm{Ne}$
collisions with collisions of nearly-spherical nuclei such as those of 
$^{16}\mathrm{O}$, as delivered by the LHC to experiments in
July 2025. The ratios of observables between the two systems are expected to be 
largely independent of final-state transport properties, and directly
access the variations in the initial condition caused by nuclear
structure differences~\cite{giacalone2024unexpectedusesbowlingpin}. 
Another, potentially superior way of imaging the structure of light nuclei is
to collide them with heavy spherical nuclei, such as in
$^{208}\mathrm{Pb}^{20}\mathrm{Ne}$ collisions~\cite{Broniowski:2014prl,Maciej:2018,Zhang:2017xda,Giacalone:2025prl,Lu:2025cni},
where, at small impact parameter,  the overlap region provides a direct snapshot of the shape of the light nucleus.

The LHCb Upgrade I detector is a single-arm forward spectrometer covering the pseudorapidity range 
$2 < \eta < 5$, described in detail in Ref.~\cite{LHCb-DP-2022-002}.
The detector elements that are particularly relevant to the analysis described in this note include:
 a silicon-pixel vertex detector surrounding the beam-beam interaction
region, a high-precision tracking system that provides a measurement of the momentum, $p$, of charged
particles, and an electromagnetic calorimeter.
LHCb has the unique capability, among the LHC experiments, to also record fixed-target collisions 
between the LHC beams and a gas target provided by the SMOG2
system~\cite{LHCb-TDR-020, LHCb-DP-2024-002}. Most of the beam-gas collisions occur
within the SMOG2 gas storage cell, located $44 \pm 10\cm$
upstream of the nominal beam-beam interaction region. During the LHC Run~3, fixed-target collisions with various gas-target species are routinely being
acquired, simultaneously with beam-beam collisions.

This note presents the first results from flow studies 
in $^{208}\mathrm{Pb}\mathrm{Ne}$ and $^{208}\mathrm{Pb}\mathrm{Ar}$
collisions at $\sqsnn = 70.9\gev$,
acquired during the LHC lead run in November 2024.
The measurement of the flow coefficients \vtwotwo and \vthreetwo, obtained with
the cumulant method~\cite{PhysRevC.63.054906,PhysRevC.64.054901,PhysRevC.83.044913}, is reported as a function of centrality. While injection of oxygen has not been possible so far for operational
reasons, the argon gas is chosen as a reference to the study of the highly
deformed $^{20}$Ne nuclear shape, since, though having a
larger size, it is expected to exhibit a nearly 
spherical shape similar to the $^{16}\mathrm{O}$ nucleus~\cite{QM25_CHUN_new}. 
The injected gas targets have natural isotopic composition, implying
that the fraction of  $^{20}\mathrm{Ne}$ nuclei in the neon target  is
90.48\%, with $^{22}\mathrm{Ne}$ (9.25\%) and $^{21}\mathrm{Ne}$
(0.27\%) isotopes also present. The abundance of
$^{40}\mathrm{Ar}$ in the argon target is 99.60\%.

The note is organised as follows: in Sec.~\ref{sec:method}, the
cumulant method is introduced and the observables are defined;
in Sec.~\ref{sec:selection}, the online and offline selection of the
data, and the weighting method used to take into account the track
reconstruction inefficiency are discussed;  in Sec.~\ref{sec:PbPb}, 
the cumulant method is validated on a sample of PbPb collisions
acquired with the Run~1 and 2 \lhcb detector in 2018, and for which 
the flow coefficients \vtwo and \vthree are measured from
two-particle correlations~\cite{LHCb-PAPER-2023-031}; 
in Sec.~\ref{sec:results},  the systematic
uncertainties of the flow-coefficient  measurements are studied and
the results presented. 
In Sec.~\ref{sec:closing}, conclusions are drawn and future outlook is given.

%%%%%%%%%%%%%%%%%%%%%%%%%%%%%%%%%%%%%%%%%%%%%%%%%%%%%%%%%%%%%%%%%%%%%%%%%%%%%%%%%%%%%%%%%%%%%%%%%
\section{The cumulant method}
\label{sec:method}
Collective flow refers to the correlated and anisotropic expansion of matter created in high-energy heavy-ion collisions. In such collisions, the initial geometric asymmetries of the overlapping nuclear regions and the subsequent pressure gradients drive the produced particles to exhibit preferred directions in their momentum distributions~\cite{Ollitrault:1992prd}. This behaviour is typically quantified by expanding the azimuthal distribution of emitted particles into Fourier harmonics~\cite{Voloshin:1996}
\begin{equation}
    \label{eq:vn_expansion}
    f(\phi) = \frac{1}{2\pi} 
    \Bigl\{1 + 2 \sum_{n=1}^{\infty} 
        v_n \cos\bigl[n(\phi - \Psi_n)\bigr]
    \Bigr\},
\end{equation}
where $\phi$ denotes the azimuthal angle and $\Psi_{n}$ the $n$th-order symmetry plane. The most commonly analysed flow coefficients $v_n$ are the first few harmonics: $v_{1}$, usually referred to as \textit{directed flow}, \vtwo, \textit{elliptic flow}, and \vthree, \textit{triangular flow}. By exploiting the orthogonality of the cosine functions in a Fourier series, each flow coefficient $v_{n}$ can be isolated via the integrals over $\phi$
\begin{equation}
    \label{eq:def_vn}
    v_n = \langle \cos\left[ n\left( \phi - \Psi_n \right) \right]\rangle,
\end{equation}
where the angular brackets denote the average over all particles in an event. 

A variety of experimental methods to calculate the $v_n$ observables are available. In this analysis, multiparticle correlations are built to cancel the dependence on the reaction plane, as proposed in Ref.~\cite{PhysRevC.63.054906,PhysRevC.64.054901,PhysRevC.83.044913}. 
A single-event $m$-particle correlation in harmonics $n_1,n_2,...,n_m$ is expressed as
\begin{equation}
    \label{eq:def_m_particle_corre}
    \langle m \rangle_{n_1,n_2,...,n_m} = \langle e^{i(n_1\phi_1+n_2\phi_2+...+n_m\phi_m)} \rangle,
\end{equation}
where the angular brackets represents an average of all $m$-particle combinations of the overall $M$ particles created in a collision event. Then, a second average over all events is computed to obtain an event-averaged correlation, generally denoted as
\begin{equation}
    \label{eq:def_m_particle_corre_average}
    \langle\langle m \rangle\rangle_{n_1,n_2,...,n_m} = \langle\langle e^{i(n_1\phi_1+n_2\phi_2+...+n_m\phi_m)} \rangle\rangle.
\end{equation}
To avoid an explicit nested loop, which is computationally expensive, the  $Q$-cumulant method~\cite{PhysRevC.83.044913} was developed and then modified in the so-called generic framework~\cite{Generic-framework}, defining the $Q$-vector $Q_{n,l}$ for the harmonic order $n$ as
\begin{equation}
    \label{eq:def_Q_n_l}
    Q_{n,l} \equiv \sum_{k=1}^M w_{k}^l e^{in\phi_{k}}.
\end{equation}
Here, each particle contributes with a weight $w$, which corrects for detector acceptance and efficiency effects in the track reconstruction, so that the measured correlations better reflect the true underlying physics.  
The superscript $l$ is an integer between 1 and $m$, specifying the power of the weight associated with the $m$-particle correlation. Using these weighted $Q$-vectors, the $m$-particle correlation defined in Eq.~\ref{eq:def_m_particle_corre} can be re-expressed in a compact form by combining $Q_{n,l}$ with different $(n,l)$ inputs, thus avoiding the direct evaluation of all particle combinations~\cite{Generic-framework}. In particular, for the $n$th-order anisotropic flow coefficient measurement presented in this note, the harmonic numbers are taken as $n_1 = n_2 = \cdots =n_m=n$. The cumulants $c_{n}\{m\}$, $m$ being the number of correlated particles, are obtained by subtracting the contributions of lower-order correlations as
\begin{equation}
\begin{split}
    \label{eq:def_c_n_m}
    c_n\{2\} = \,& \langle\langle 2 \rangle\rangle_{n,-n},  \\
    c_n\{4\} = \,& \langle\langle 4 \rangle\rangle_{n,n,-n,-n} - 2 \cdot \langle\langle 2 \rangle\rangle^{2}_{n,-n}.
\end{split}
\end{equation}
The flow coefficients with two or four correlated particles can finally be computed from cumulants as
\begin{equation}
\begin{split}
\label{measure_v_n_m}
    v_n\{2\} &= \, \sqrt{c_n\{2\}},  \\
    v_n\{4\} &= \, \sqrt[4]{-c_n\{4\}}.
\end{split}
\end{equation}
For the same $v_n$, the two estimators above are subject to different biases. While higher-order correlations have larger statistical uncertainties, they have lower sensitivity to nonflow effects and higher sensitivity to eccentricity fluctuations~\cite{Borghini:2001zr,PhysRevC.84.024911}. To reduce nonflow contributions, \ie dijets and other short-range correlations that significantly distort the multiparticle cumulant measurements, the subevent cumulant method is used~\cite{Subevent-1,Subevent-2}. This consists of dividing the considered $\eta$ range into two or more disjoint regions, from which particles in correlation are taken. In this analysis, two subevent regions with a $|\Delta \eta| = 1$ excluded region are considered.

Differential $v_n$ flow measurement as a function of the transverse momentum, \pt, are also presented in this work. Unlike the integrated case described above, tracks within a single event are categorized into several \pt intervals. To reduce statistical uncertainties, each differential measurement is performed by selecting one particle from the \pt interval of interest and others from a predefined broad reference range. The formula in Eq.~\ref{eq:def_m_particle_corre} is then modified as
\begin{align}
    \label{eq:differential-correlations}
    \langle m'\rangle_{n_{1},n_{2},...,n_{m}} = \,&
    \langle e^{i(n_{1}\psi_{k_{1}}' + n_{2}\phi_{k_{2}} + ... + n_{m}\phi_{k_{m}})} \rangle,
\end{align}
where $\psi_{k_{1}}'$ indicates that the particle is taken from the \pt interval of interest, while the remaining particles are selected from the reference window.
The two- and four-particle differential cumulants are defined as
\begin{equation}
    \begin{split}
    \label{eq:diff_flow_cumulant}
    d_{n}\{2\} = \, & \langle\langle 2' \rangle\rangle , \\
    d_{n}\{4\} = \, & \langle\langle 4' \rangle\rangle - 2 \cdot \langle\langle 2' \rangle\rangle\langle\langle 2 \rangle\rangle.
    \end{split}
\end{equation}
The corresponding differential flow coefficients are finally obtained by normalising the differential cumulants with the reference results given in Eq.~\ref{eq:def_c_n_m},
\begin{equation}
    \begin{split}
    \label{eq:diff_flow_coefficient}
    v'_{n}\left\{2\right\} = \,& \frac{d_{n}\left\{2\right\}}{\sqrt{c_{n}\left\{2\right\}}} , \\
    v'_{n}\left\{4\right\} = \,& -\frac{d_{n}\left\{4\right\}}{(-c_{n}\left\{4\right\})^{\frac{3}{4}}}.
    \end{split}
\end{equation}

%%%%%%%%%%%%%%%%%%%%%%%%%%%%%%%%%%%%%%%%%%%%%%%%%%%%%%%%%%%%%%%%%%%%%%%%%%%%%%%%%%%%%%%%%%%%%%%%%
\section{Fixed-target PbA collisions at \lhcb}
\label{sec:selection}

The main datasets used for this analysis consist of PbNe and PbAr collisions at \mbox{$\sqsnn = 70.9$\gev},  corresponding to an integrated luminosity of about 0.06 and 1.7\invnb, respectively. Neon and argon gases were alternatively injected into the SMOG2 cell, and data acquired, reconstructed and selected concurrently with the collisions of the Pb beams. 

Simulated samples are required to model the effects of the detector acceptance and the
imposed selection requirements. In the simulation, PbNe and PbAr collisions are generated using \textsc{EPOS}~\cite{EPOS-LHC}. Decays of unstable particles are described by \evtgen~\cite{Lange:2001uf}, in which final-state radiation is generated using \photos~\cite{davidson2015photos}. The interaction of the generated particles with the detector, and its
response, are implemented using the \geant toolkit~\cite{Allison:2006ve, *Agostinelli:2002hh} as described in Ref.~\cite{LHCb-PROC-2011-006}.

The online selection of beam-gas collision events requires a minimum-bias condition that is fully efficient for the hadronic interactions in the centrality range 
of interest for this study. It requires that a collision primary vertex (PV) is reconstructed within the SMOG2 cell and that the total energy measured in the calorimeter \EcalE exceeds $94$\gev. The minor contribution of high-occupancy events where the tracking detectors are saturated due to background PbPb collisions, as described in the following, are discarded.
Since the measurement is not expected to be limited by the statistical uncertainty, for the large PbAr sample only a randomly selected fraction of 2\% 
of the events passing the online requirements were recorded. The analysis is based on 102 (70) million PbNe (PbAr) events.

\begin{figure}[tb]
    \centering
    \begin{minipage}[t]{0.49\linewidth}
        \centering
        \includegraphics[width=\linewidth]{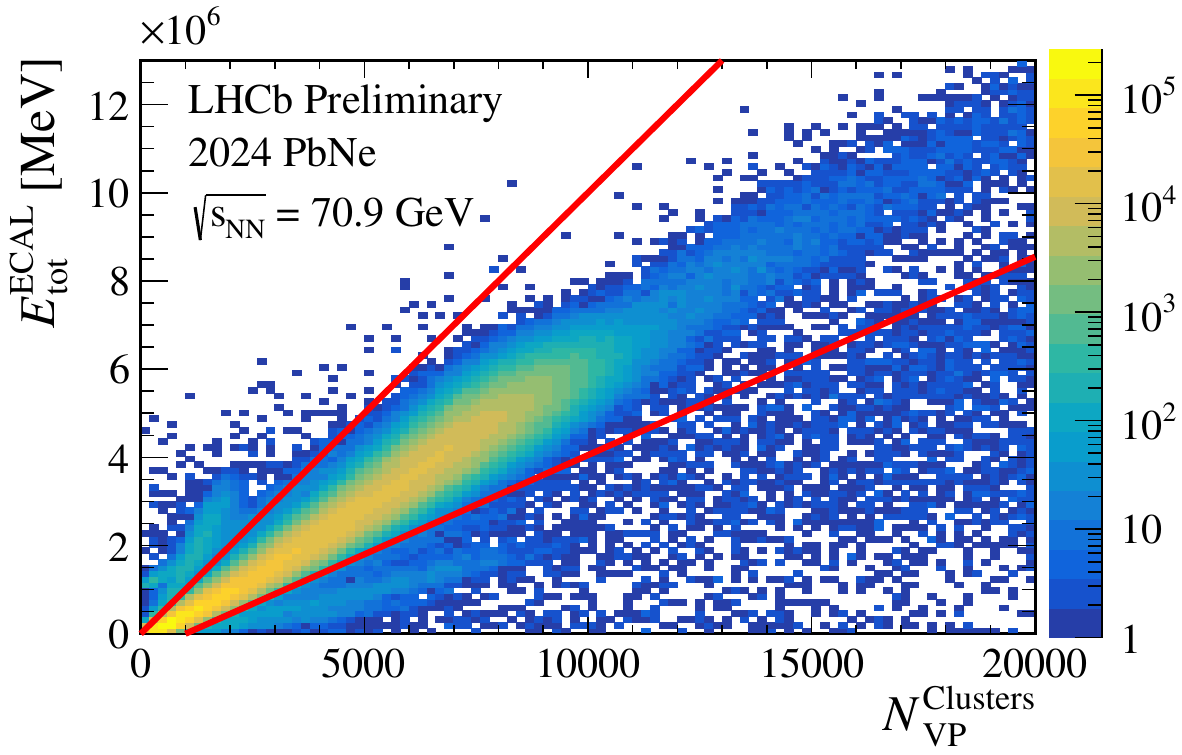}
    \end{minipage}
    % \hfill
    \begin{minipage}[t]{0.49\linewidth}
      \centering
      \includegraphics[width=\linewidth]{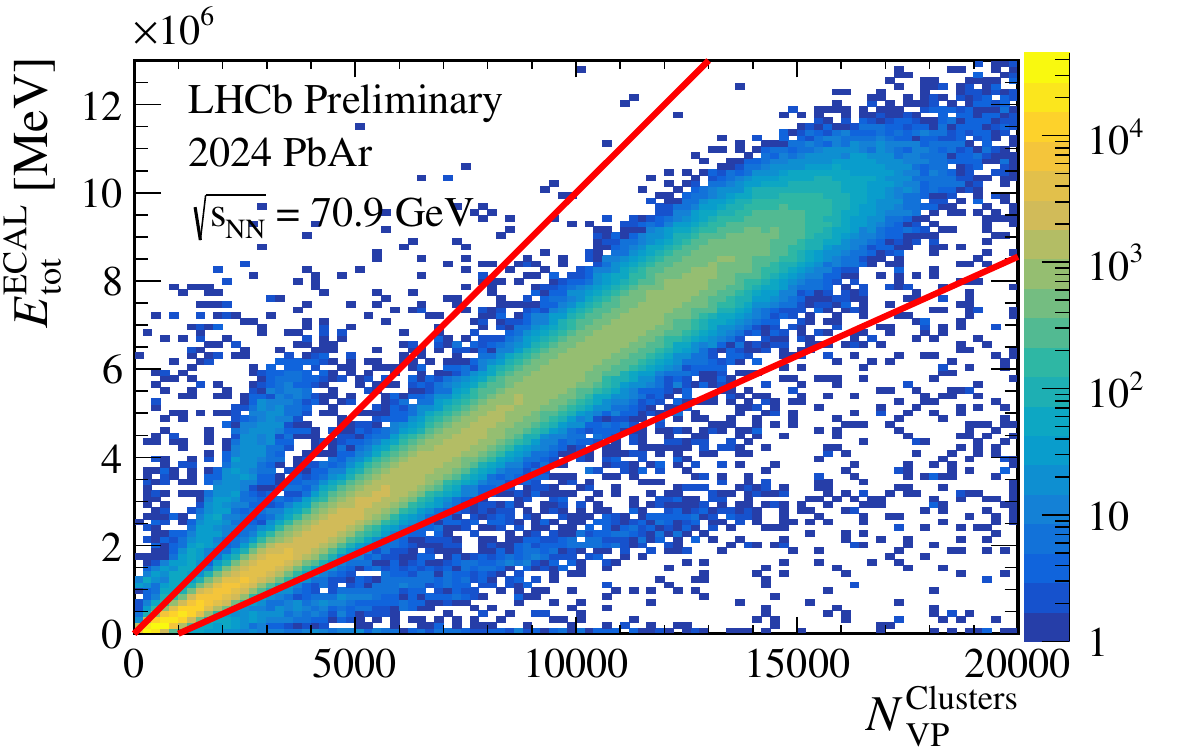}
    \end{minipage}
    \begin{minipage}[t]{0.49\linewidth}
        \centering
        \includegraphics[width=\linewidth]{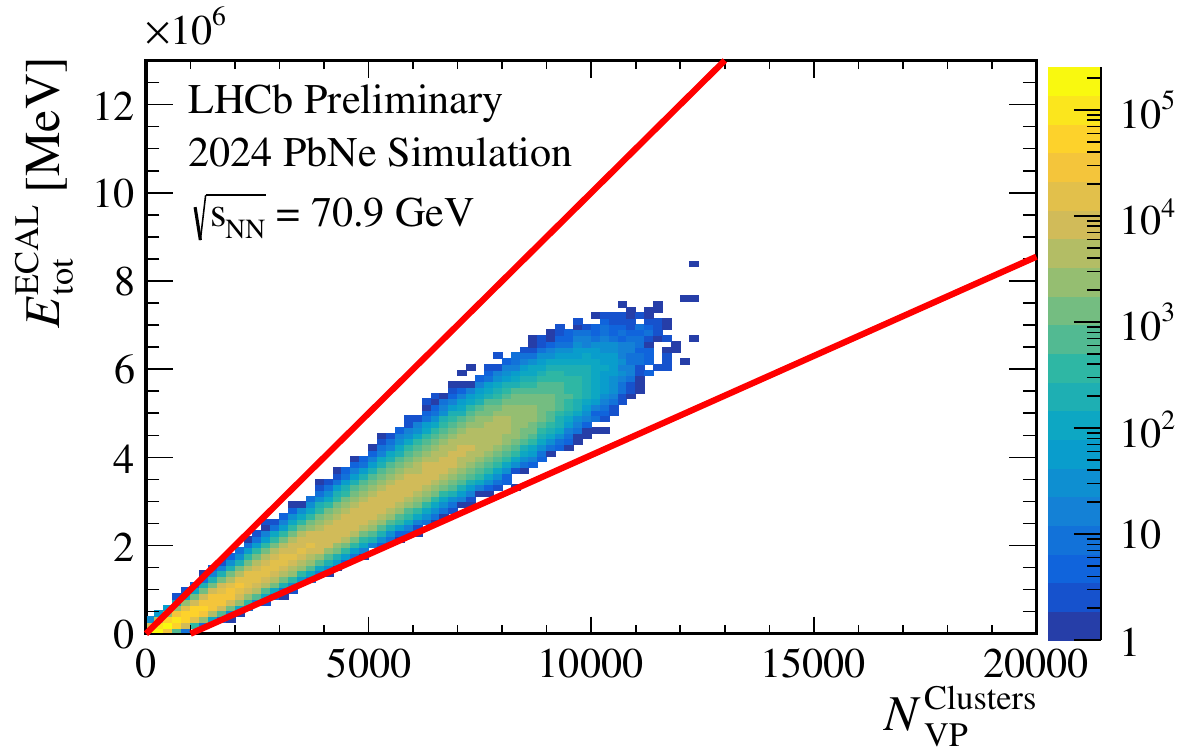}
    \end{minipage}
    % \hfill
    \begin{minipage}[t]{0.49\linewidth}
        \centering
        \includegraphics[width=\linewidth]{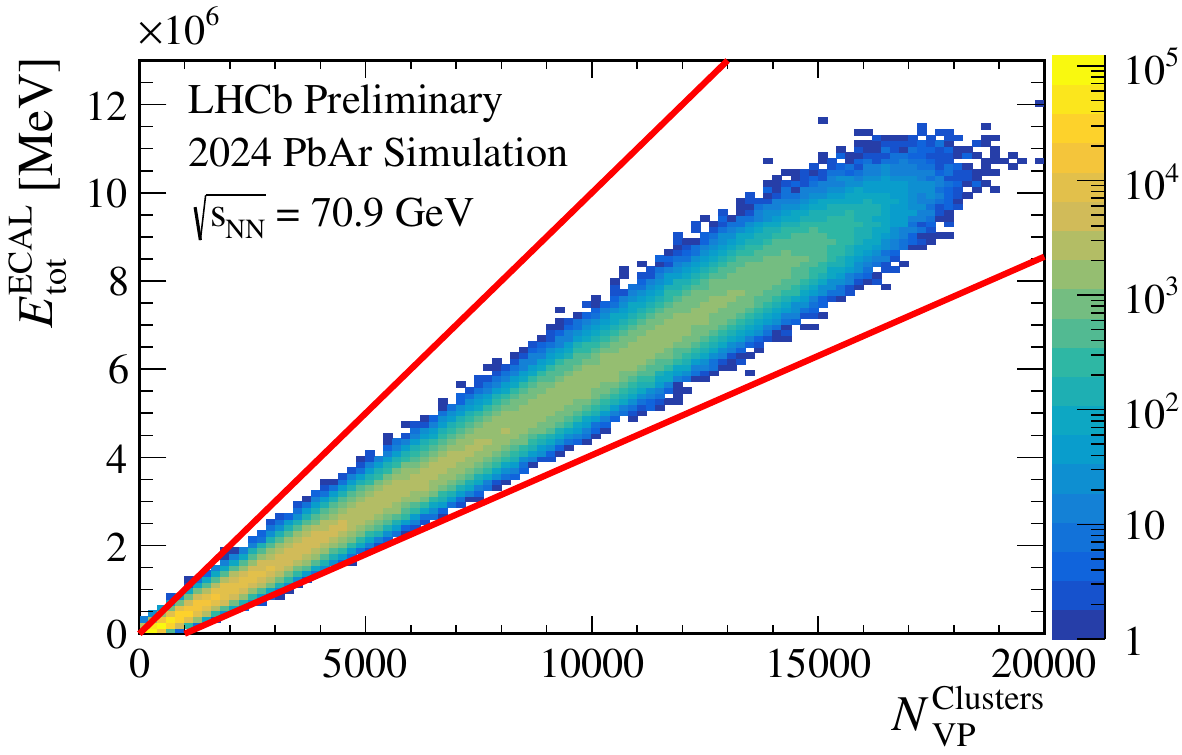}
    \end{minipage}
    \caption{Examples of two-dimensional distributions of \EcalE vs \NVPClusters in (left) PbNe and (right)~PbAr  collision for (top) data and (bottom) simulation, before any selection is applied. As detailed in the text, only events in between the two red lines, corresponding to PbA collisions in the SMOG2 cell, are considered for the analysis.}
    \label{fig:hist_2d_Ecal_nVP}
\end{figure}

In the offline analysis, additional requirements are set to suppress background events.
Figure~\ref{fig:hist_2d_Ecal_nVP} shows the correlation between \EcalE and the number of hit clusters
reconstructed in the vertex detector \NVPClusters. Events are required to satisfy the relation
\mbox{$(0.45\cdot\NVPClusters - 450) < \EcalE/\gev< \NVPClusters$}, covering all hadronic interactions according to simulation.
The bands visible outside the selected range are attributed to collisions occurring outside the SMOG2 cell and 
electromagnetic interactions in ultraperipheral events.
Beams can interact with gas atoms flowing out from the cell toward the closest vacuum pumps (located about 20 m away on either side).
Collisions occurring meters away from the cell can produce high-energy particles showering through the beam pipe and reaching the detectors.
These events produce a much larger \NVPClusters value for a given energy deposit \EcalE  with respect to the 
proper SMOG2 collisions. Collisions of PbPb can also produce background events
by increasing the rate of fake PVs reconstructed within the cell, while the probability of two simultaneous PbA and PbPb collisions 
from the same beam bunch is negligible.

Multiple requirements on the reconstructed PV are imposed to suppress the background contributions from fake PVs and/or secondary vertices.
The PV position is required to be compatible with the measured beam trajectory and must be formed from at least seven tracks.
In a few percent of events, more than one PV passes these requirements and in such cases only that with the highest track multiplicity
is considered. 
As the data show evidence of cases where a physical vertex is split into two reconstructed PVs, those that are closer than 10~mm
to the selected PV are merged with it.
The simulation shows that this PV selection procedure suppresses the background from fake PVs to a negligible level. It also guarantees that tracks
from different collisions are not mixed also in the rare case, with a probability smaller than $0.1\%$, when more than one beam-gas collision occurs
inside the cell.
The offline selection requirements remove less than 2\% of the acquired
collision events.

\begin{figure}[tb]
    \centering
    \includegraphics[width=.8\linewidth]{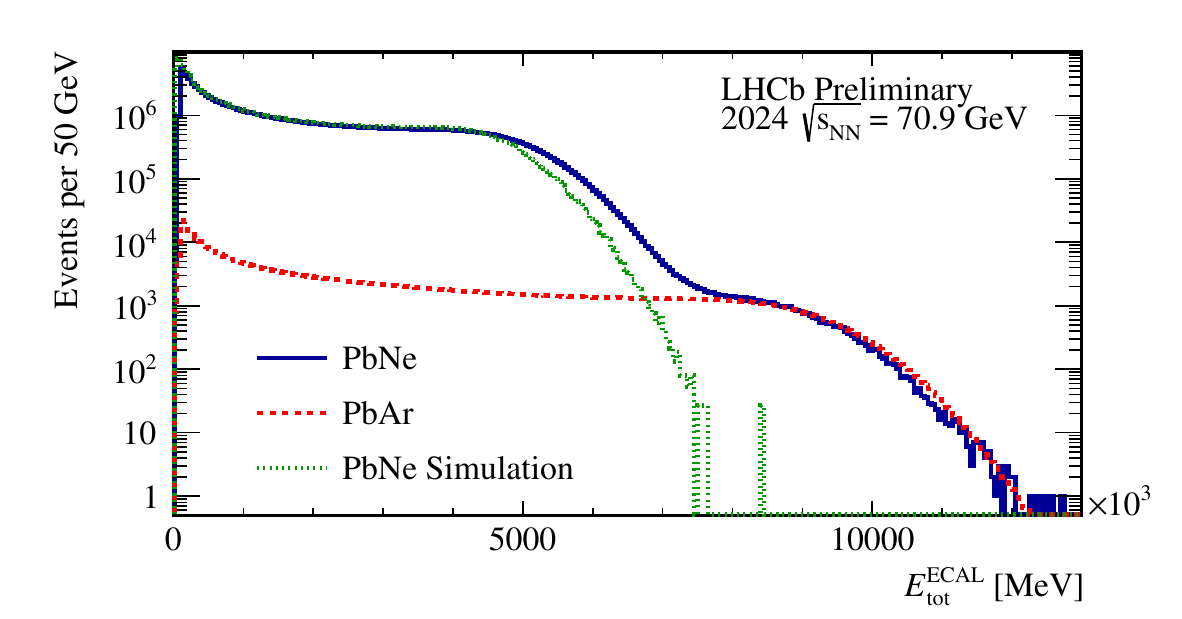}
    \caption{Comparison of the \EcalE distribution for PbNe and PbAr collisions, after normalising the latter to the interval $8 < \EcalE < 11\tev$. Here, it is assumed that  PbAr collisions also dominate in the PbNe sample, as expected from simulation. }
    \label{fig:ArgonContamination}
\end{figure}

An anomalous tail in the  \EcalE distribution for PbNe collisions is found to be due to residual contamination from argon which was injected before the neon runs.
Such contamination is indeed observed to depend on the time elapsed since the last argon injection. 
This demonstrates that, after  the purging of the SMOG2 injection system and the change of injected gas, a small amount of argon persists in the beam pipe 
with a relaxation time of a few hours, probably because of the long time needed to remove the gas from the injection capillary.
The total contamination of events from PbAr collisions in the analysed PbNe sample is estimated to be lower than 0.5\% from the tail of
the  \EcalE distribution, as illustrated in Fig.~\ref{fig:ArgonContamination}. In the  region $6 < \EcalE < 8\tev$, corresponding to the 0.5\% most central PbNe collisions,
the contamination is larger than 2\% and is considered among the systematic uncertainties. The region $\EcalE > 8\tev$, dominated by PbAr collisions, is removed from the  PbNe sample.

\begin{figure}[tb]
    \centering
    \begin{minipage}[t]{0.49\linewidth}
        \centering
         \includegraphics[width=\linewidth]{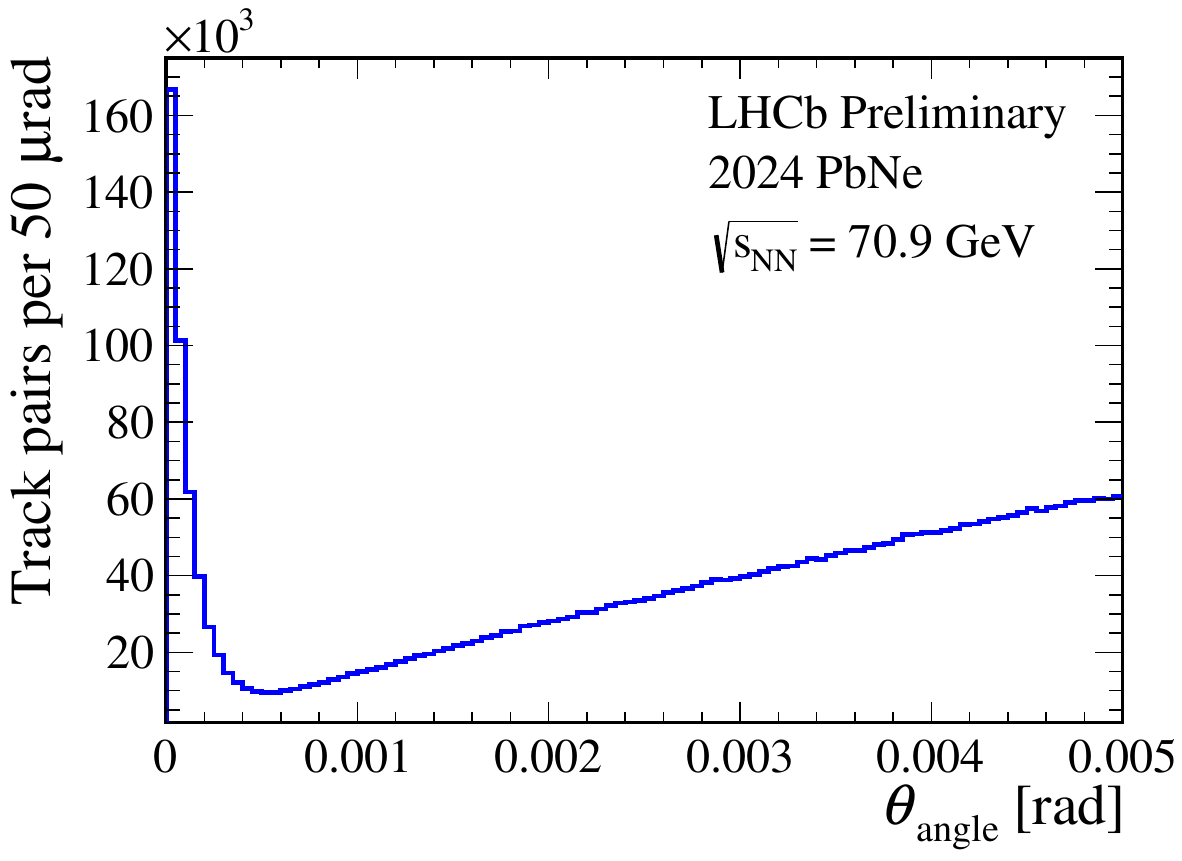}
    \end{minipage}
    % \hfill
    \begin{minipage}[t]{0.49\linewidth}
        \centering
        \includegraphics[width=\linewidth]{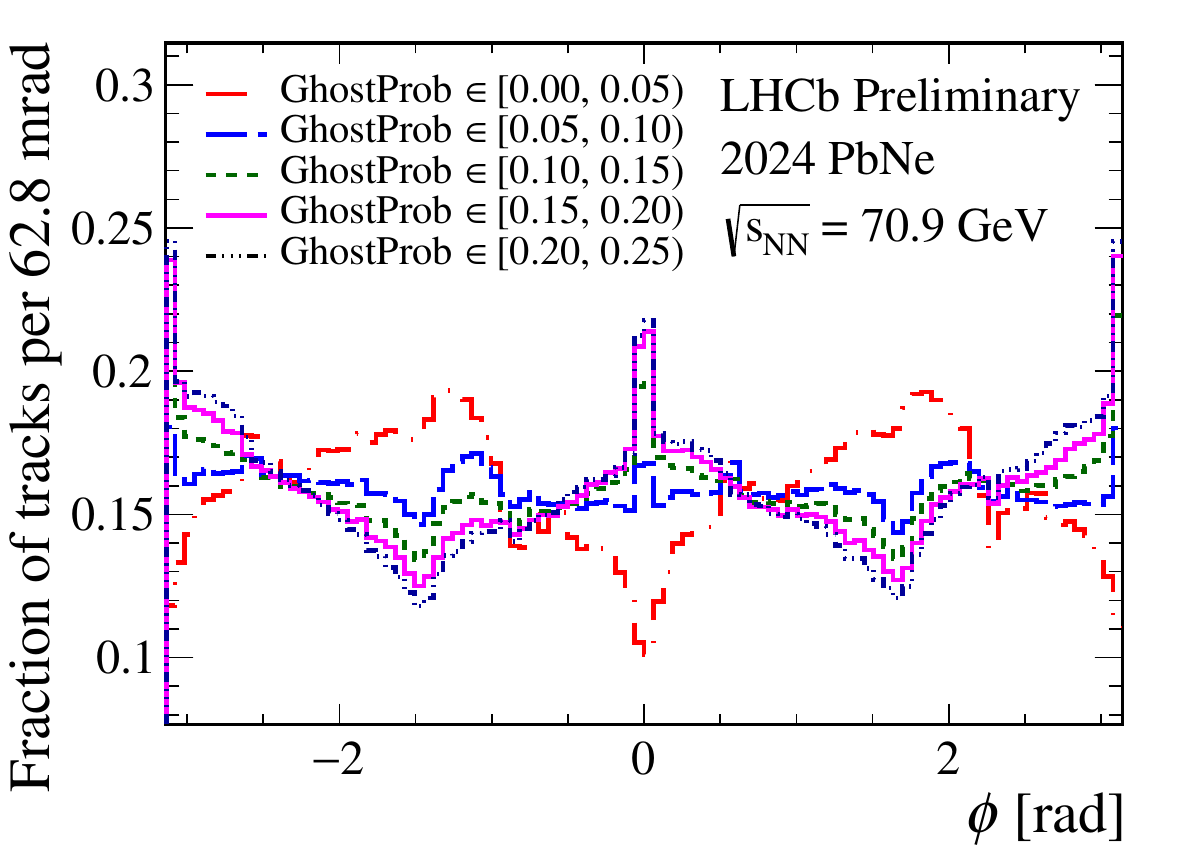}
    \end{minipage}
    \caption{Distributions prior to track selection of (left) the \clonesangle variable  and (right) 
      the azimuth angle $\phi$ in different intervals of the
      GhostProb variable for candidate tracks in PbNe collisions.} 
    \label{fig:cloneghosts}
\end{figure}

To compute the cumulant observables only from prompt particles produced in a single collision,
selected tracks are required to be compatible with originating from the selected PV. The impact parameter must be smaller than 1\mm
and the difference in \chisq of the PV reconstructed with or without the considered track 
must be lower than 9. 
Reconstructed particles are selected in the phase space corresponding to the detector acceptance.
Momentum is required to be larger than 2\gev,\footnote{Natural units are used throughout the paper.}
the transverse momentum in the $0.2<\pt<3.0\gev$ range
and the pseudorapidity in the $2.4<\eta<5.0$ range, where the lower bound is tighter than the nominal \lhcb acceptance
because of the displaced interaction region for fixed-target collisions.

Particular attention is devoted to suppressing ``clone tracks'', which originate from the same particle hits, 
and ``ghost tracks'', which result from the incorrect combination of hits from different particles in the tracking detectors. 
Both types of misreconstruction can mimic flow-like signals in high-multiplicity events.
Clones are suppressed by computing the opening angle \clonesangle between the considered track and any other track.
A neural-network-based estimator called GhostProb~\cite{DeCian:2255039} is used to identify ghost tracks.
Figure~\ref{fig:cloneghosts} shows the distributions of these two quantities. Clone tracks produce 
a peak at zero opening angle, while ghost tracks produce anomalous structures in the distribution of the azimuth angle $\phi$. 
These backgrounds are suppressed by the requirements $ \clonesangle>1\mrad$ and GhostProb $<0.1$, which are varied to
 estimate the related systematic uncertainty.

As introduced in Sec.~\ref{sec:method}, the tracking inefficiency of the
detector is taken into account by weighting 
each candidate track by the inverse of its estimated efficiency.
From studies on simulation, the efficiency is found to depend mostly
on the track kinematics ($\eta$, $\phi$, \pt) and the detector
occupancy, quantified by \NVPClusters. Distortions of the  $\phi$
distribution due to the detector efficiency produce the largest bias
on the flow coefficients. The $\phi$ modulation is corrected using data from the four-dimensional distribution of the
considered variables in the selected samples. This exploits the fact that
the $\phi$ distribution, computed with respect to the measured beam
direction, is expected to be uniform for a fully efficient
detector. Weights are computed to flatten the
$\phi$ distribution in each ($\eta$, \pt, \NVPClusters)
interval, while the dependence of the detector efficiency on the ($\eta$,
\pt, \NVPClusters) variables is corrected using the predicted track
efficiency $\epsilon_{\rm{Sim}}$ and purity $\rho_{\rm{Sim}}$ from simulation. Hence, the resulting weight is
\begin{align}
    \label{eq:phi_weight}
   w_{\phi}(\phi, \eta, \pt, \NVPClusters) =
   \dfrac{\rho_{\rm{Sim}}(\eta,\pt, \NVPClusters)}{\epsilon_{\rm{Sim}}(\eta,\pt, \NVPClusters)} \cdot 
\frac{\langle N(\delta\eta,\delta \pt, \delta \NVPClusters) \rangle}{N(\delta \phi,\delta\eta,\delta \pt, \delta \NVPClusters)},
\end{align}
where $N$ is the number of selected tracks in data for each four-dimensional interval and $\langle N \rangle$ its average over the $\phi$ variable. 
One hundred intervals are used for the $\phi$ and \NVPClusters variables, and ten for $\eta$ and \pt,
their size being chosen to obtain roughly equal statistics in each.

%%%%%%%%%%%%%%%%%%%%%%%%%%%%%%%%%%%%%%%%%%%%%%%%%%%%%%%%%%%%%%%%%%%%%%%%%%%%%%%%%%%%%%%%%%%%%%%%%
\section{Validation of the method with PbPb collisions}
\label{sec:PbPb}
\begin{figure}[tb]
    \centering
    \includegraphics[width=0.49\linewidth]{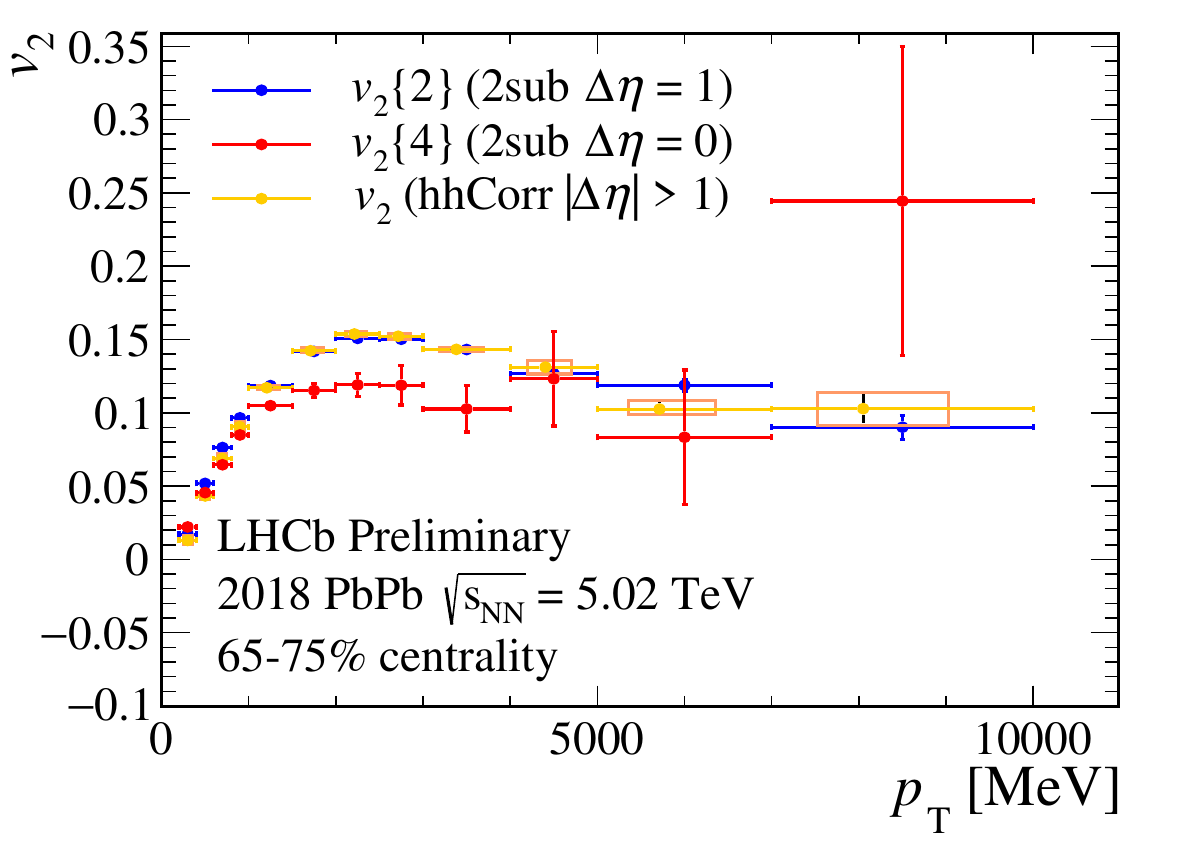}
    \includegraphics[width=0.49\linewidth]{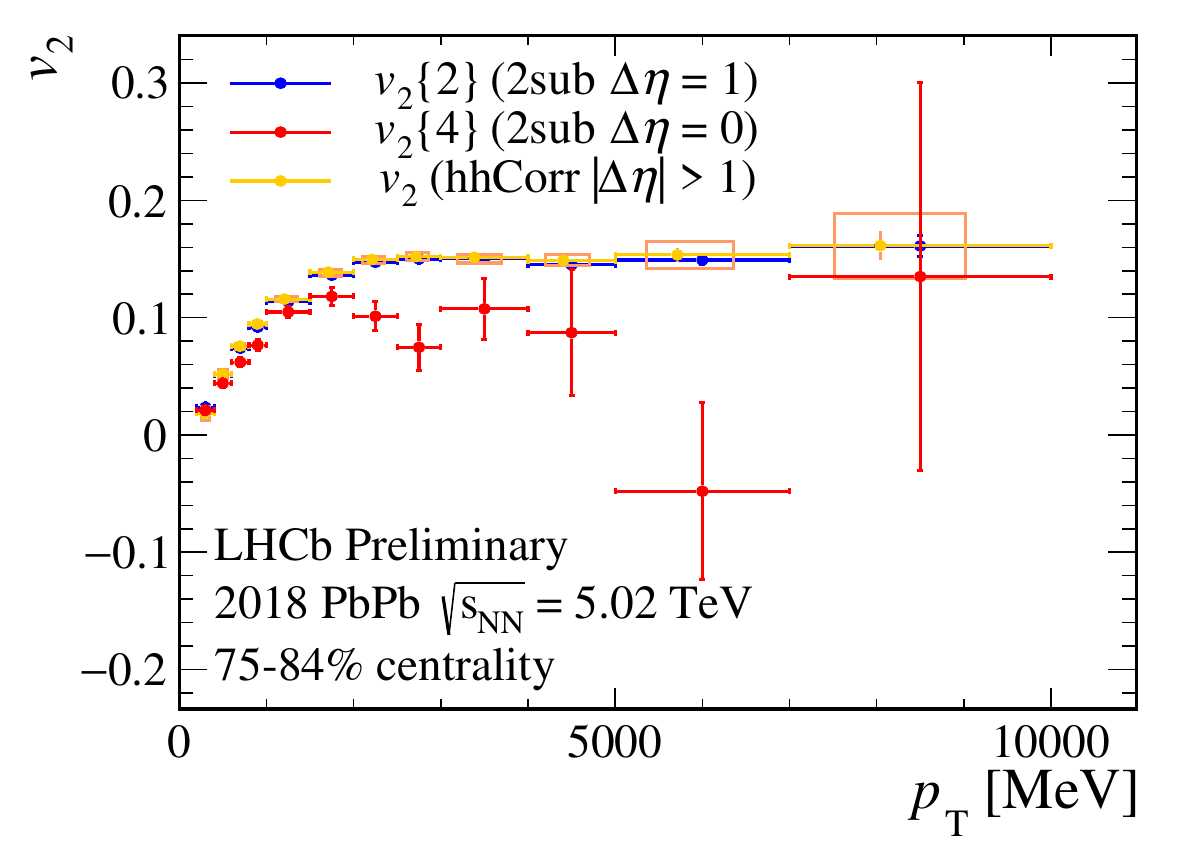}
    \includegraphics[width=0.49\linewidth]{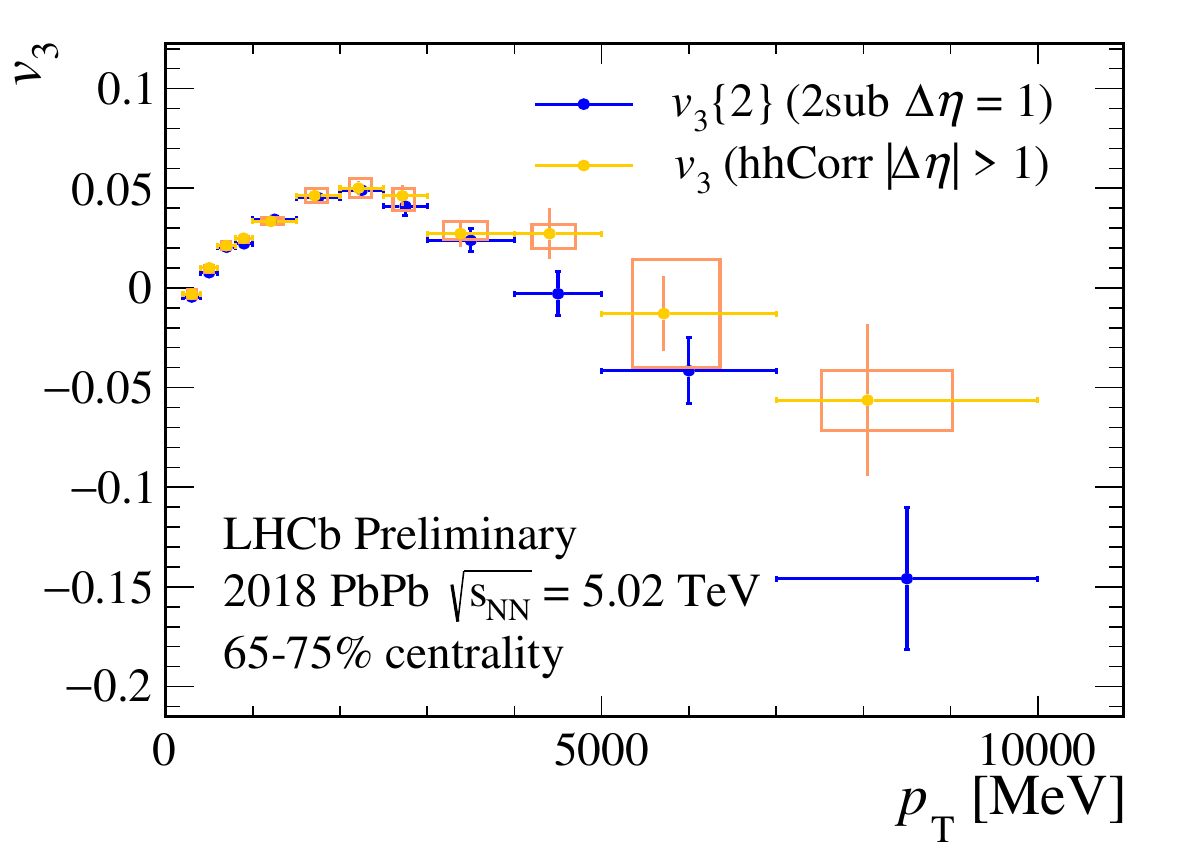}
    \includegraphics[width=0.49\linewidth]{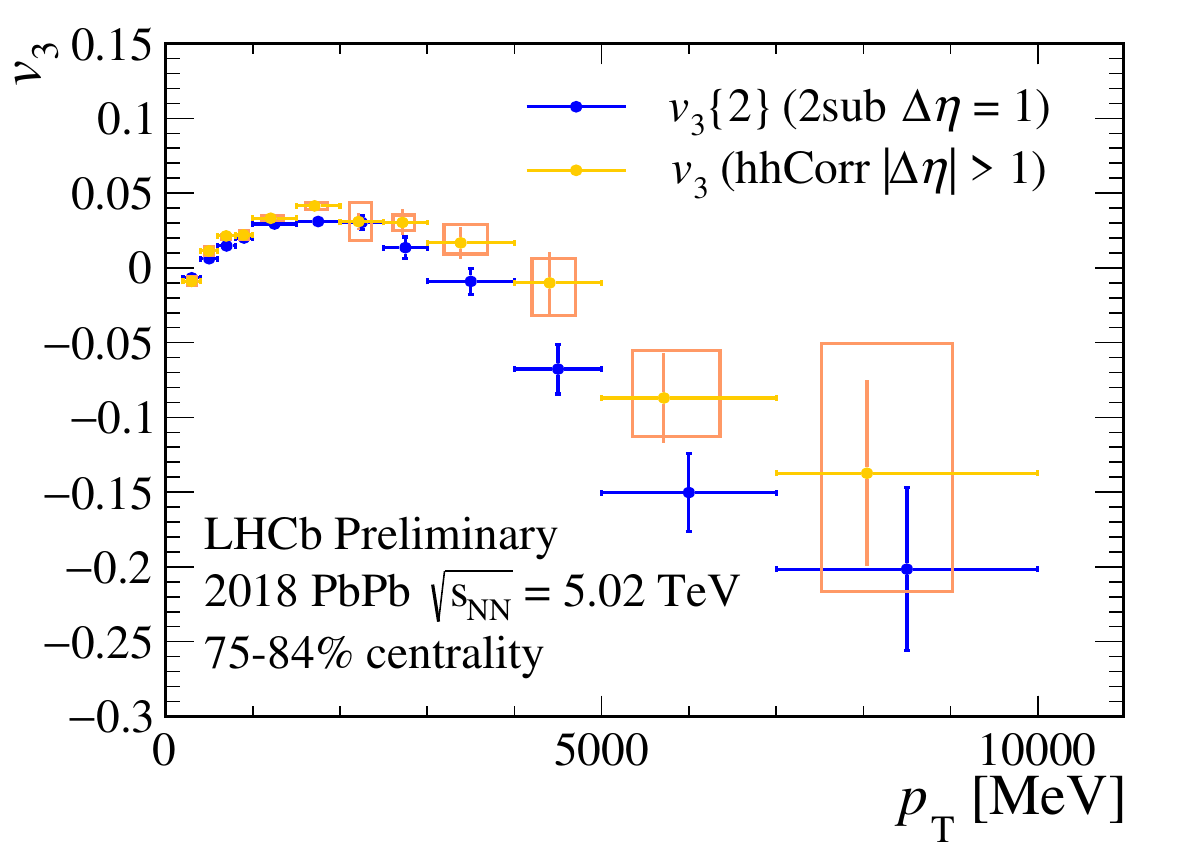}
    \caption{Differential (top) \vtwotwo and \vtwofour and (bottom) \vthreetwo flow coefficients as a function of \pt for $\sqsnn = 5.02\tev$ PbPb collisions in the (left) 65--75\% and (right) 75--84\% centrality classes. The \vtwotwo (\vtwofour) results, in blue (red), are computed with the multiparticle cumulant method with two subevents and a $|\Delta \eta|=1.0$ ($|\Delta \eta|=0$) excluded region. The \vtwo and \vthree results in orange are obtained from the two-particle correlation method~\cite{LHCb-PAPER-2023-031}.}
    \label{fig:PbPb_validation}
\end{figure}

The cumulant method discussed in Sec.~\ref{sec:method} to calculate the $v_n$ flow coefficients, the selection, detector acceptance and efficiency-correction procedures described in Sec.~\ref{sec:selection} are first validated on PbPb data collected by \lhcb in 2018 at $\sqsnn = 5.02\tev$, corresponding to an integrated luminosity of 213.7\invub. For this dataset, the \vtwo and \vthree coefficients have already been measured by \lhcb~\cite{LHCb-PAPER-2023-031}, by exploiting the two-particle correlation method.
Due to the saturation in the LHCb Run~1~and~2 tracking detectors for PbPb collisions with centralities above 60\%, and to remove contamination from ultraperipheral collisions, only the two centrality classes 65--75\% and 75--84\% were considered. The resulting \vtwo and \vthree are shown in Fig.~\ref{fig:PbPb_validation} as a function of \pt.

The same flow coefficients are computed in this work  by exploiting
the multiparticle cumulant method, following a similar procedure to that
summarised in Sec.~\ref{sec:selection}. The same event-
and track-level selections as in Ref.~\cite{LHCb-PAPER-2023-031} are
applied, and efficiency effects are corrected by applying the
$\phi$-flattening procedure discussed in Sec.~\ref{sec:selection}. The cumulants are first computed for the integrated sample over the considered $0.2 < \pt < 3.0\gev$ range, initially applying the subevent method with the $|\Delta\eta| = 1$ region excluded. The size of the excluded region is then varied to 0.6 and 0.0. Being more affected by nonflow contributions, the  corresponding \vtwotwo values are found to change, while consistency is observed for \vtwofour. To minimise the statistical uncertainty on the measured \vtwofour values, no region of $\eta$ is excluded in their determination. For each \pt interval considered in Ref.~\cite{LHCb-PAPER-2023-031}, multiparticle cumulants are then computed by considering particles with $\eta$ in the other subevent, but with $0.2 < \pt < 3.0\gev$. The results for the \vtwotwo, \vtwofour and \vthreetwo flow coefficients are shown in Fig.~\ref{fig:PbPb_validation}, compared to the previously obtained results with the two-particle correlation method. Good consistency is observed between the two measurements of both \vtwo and \vthree, confirming the validity of the cumulant method in its first use at LHCb in this analysis. 
Smaller values are found for \vtwofour, consistent with previous
observations at RHIC~\cite{STAR:2002hbo} and LHC~\cite{ALICE_flow_PbPb_v24,CMS_flow_PbPb_v24}, and as expected from lower nonflow effects and/or from the larger smearing
effect due to eccentricity fluctuations.

Another set of weights is computed using  Eq.~\ref{eq:phi_weight}
to correct for efficiency effects also depending  on \pt, $\eta$, and
multiplicity, quantities not taken into account in the central values of the
results in Fig.~\ref{fig:PbPb_validation}.
After recomputing the flow coefficients with this alternative
weighting strategy, consistency within the statistical uncertainties 
of the results in Fig.~\ref{fig:PbPb_validation} is found.

%%%%%%%%%%%%%%%%%%%%%%%%%%%%%%%%%%%%%%%%%%%%%%%%%%%%%%%%%%%%%%%%%%%%%%%%%%%%%%%%%%%%%%%%%%%%%%%%%
\section{Results for PbNe and PbAr collisions}
\label{sec:results}

\begin{figure}[tb]
    \centering
\includegraphics[width=0.8\linewidth]{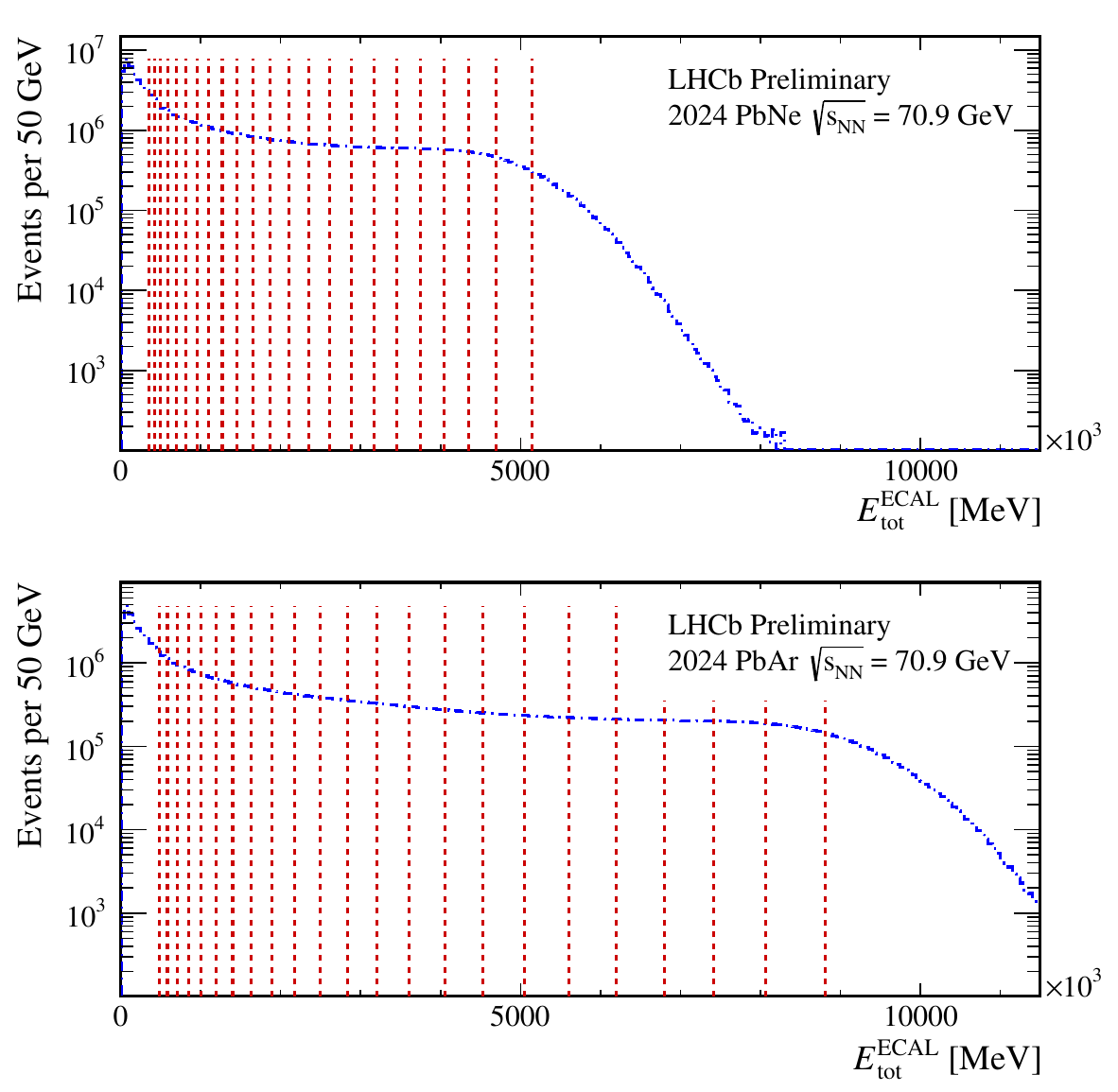}
    \caption{Assumed \EcalE spectrum for (top) PbNe and (bottom) PbAr
      collisions for the definition of the centrality proxy \cproxy. 
       Quantile boundaries for the centrality intervals of interest for this analysis, namely from 0 to 69\% in steps of 3\%, are shown.}
    \label{fig:cproxy}
\end{figure}

The cumulant method, validated in the previous section, is applied to the fixed-target PbNe and PbAr collision data
after the event and track selection described in Sec.~\ref{sec:selection}.
A preliminary estimator of the centrality in these events is provided
by the complementary quantile function of the ECAL energy distribution
$\cproxy \equiv 1-q(\EcalE)$. 
After correcting for the argon contamination in the PbNe sample (see Fig.~\ref{fig:ArgonContamination}),
the distribution of this variable is found to be in good agreement with that predicted  by simulation,
based on the Glauber-based \textsc{EPOS} model~\cite{Pierog:2013ria}, for all generated hadronic
events. The distribution measured in data is used for $\EcalE>500\gev$, 
 while for peripheral events at lower energy, affected by
reconstruction inefficiency, the simulated distribution is 
used, after normalising it to the distribution observed in data in the
range 1--6\tev (1--8\tev) for PbNe (PbAr) collisions. 
The resulting assumed \EcalE distributions are shown in
Fig.~\ref{fig:cproxy}.
The cumulants are calculated for values of estimated centrality between 0 and 69\%.

\begin{figure}[tb]
    \centering
    \begin{minipage}[t]{\linewidth}
        \centering
        \includegraphics[width=.49\linewidth]{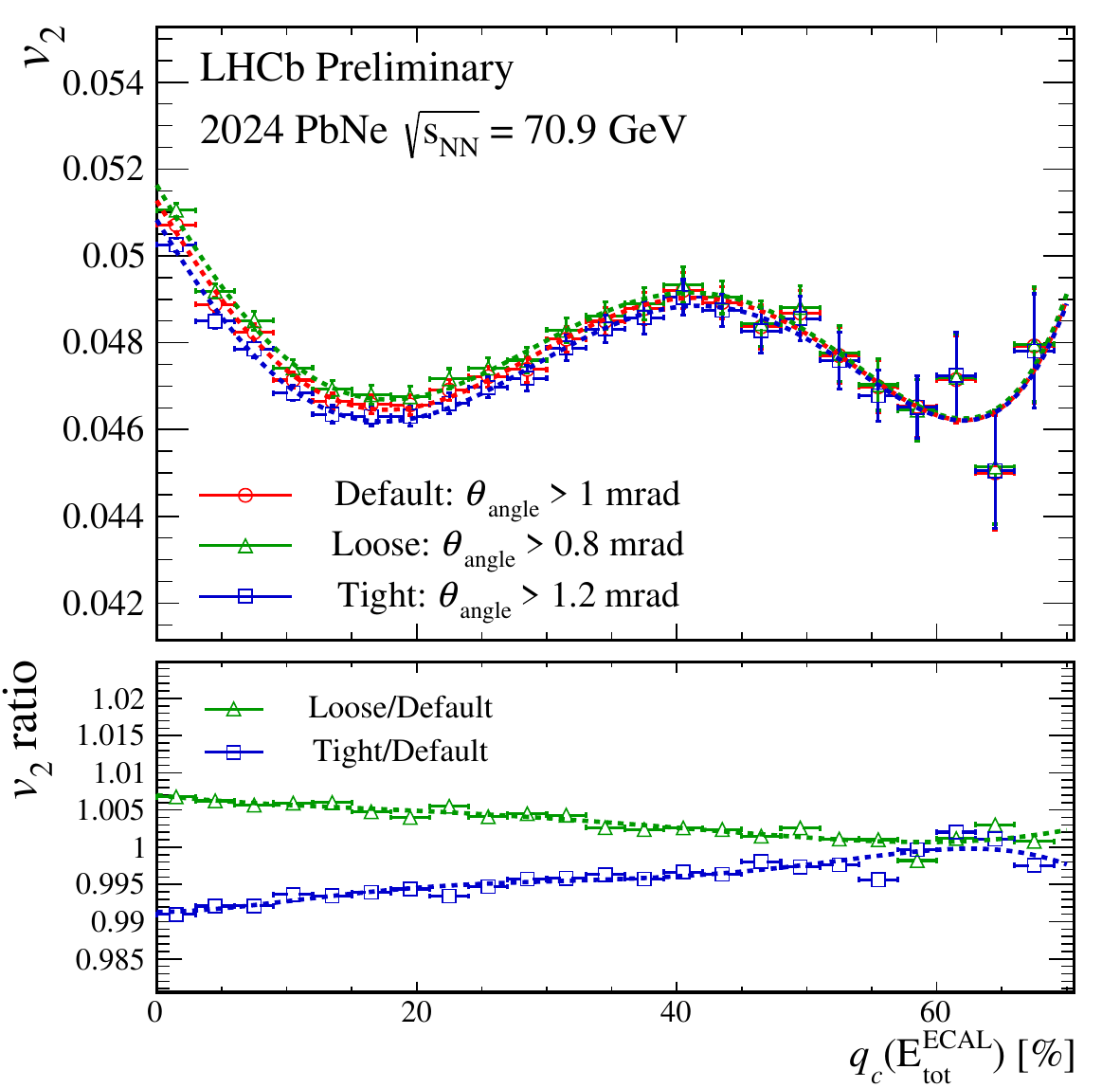}
        \includegraphics[width=.49\linewidth]{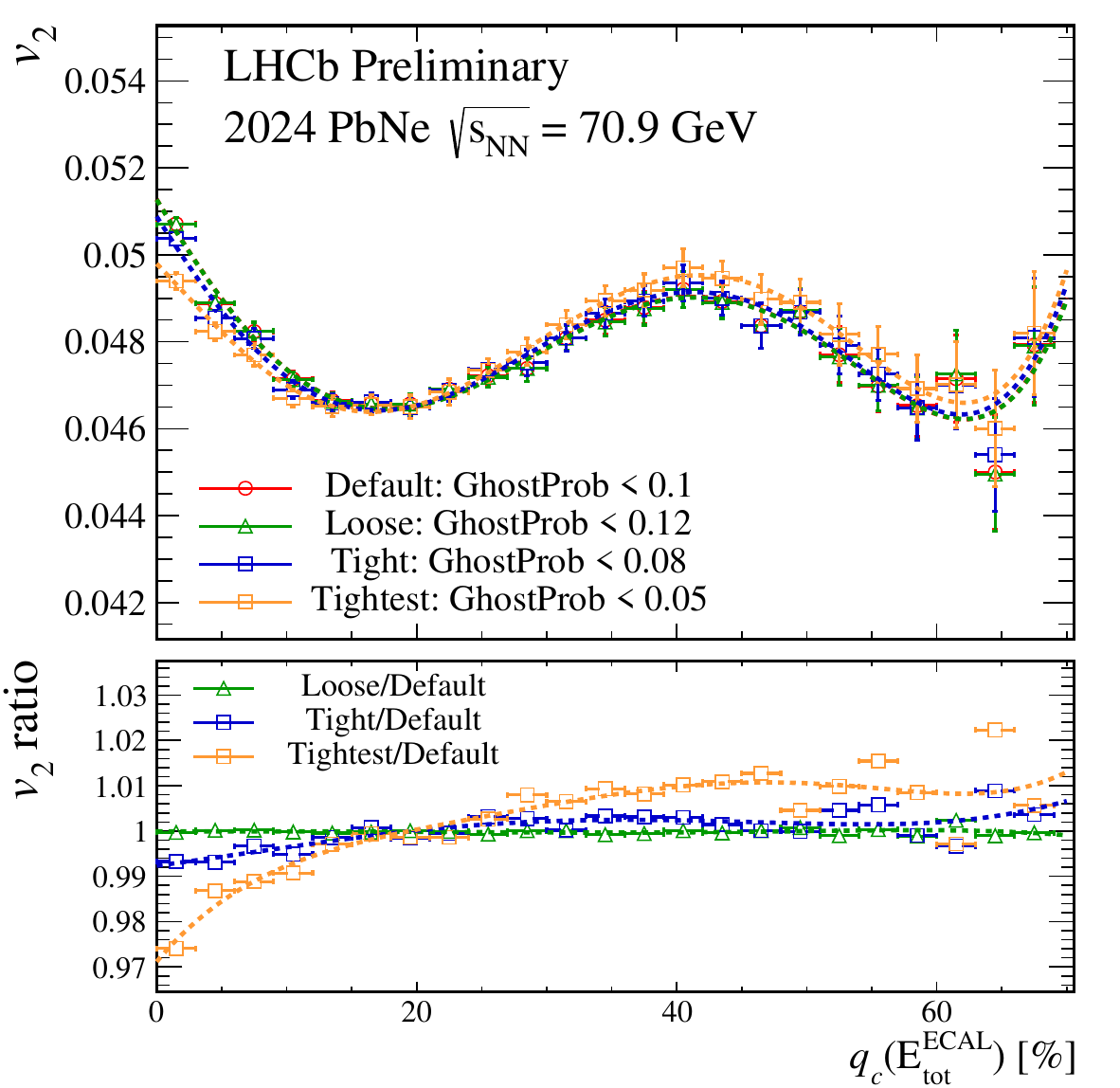}
    \end{minipage}
    \caption{Estimation of the systematic uncertainty due to (left) clone tracks and (right) ghost tracks for PbNe collisions.
    Top: \vtwotwo coefficients vs. \cproxy with different selection criteria, shown with data points and their polynomial fits. Bottom: ratios of \vtwotwo data points and fitted \cproxy dependences to the nominal results, used to evaluate the systematic uncertainties.}
    \label{fig:syst}
\end{figure}

 Several sources of systematic uncertainties  on the $v_n$
 measurements are considered, as summarised in
 Table~\ref{tab:systematics}. Possible sources of bias could result
 from residual contributions of clone, ghost or secondary tracks, as
 discussed in Sec.~\ref{sec:selection}. In each
 case, the relevant selection requirement is tightened by an amount
 chosen from the distributions of 
 the related variables in PbNe and PbAr data. 
 To smooth out statistical fluctuations, the resulting variations of the flow
 coefficients are evaluated after modelling their dependence on \cproxy, as illustrated
 by the examples shown in Fig.~\ref{fig:syst}.
 Fifth-order polynomial functions are used, as they are found to correctly
 reproduce the data points. 
 The absolute values of the resulting variations of the flow
 coefficients are assigned as systematic uncertainties.

 No significant variation is observed by tightening the impact parameter requirement on
 either the \vtwotwo and \vthreetwo result, indicating that
 secondary particles are not affecting the results. For clone and
 ghost tracks, variations up to 3\% (14\%) of the \vtwotwo (\vthreetwo) values 
 are observed for the most central collisions, and are larger for PbAr than for the PbNe data.  The possible effect of 
 the argon contamination in the neon sample is quantified by varying 
 the maximum allowed \EcalE from 8 to 6 \tev.
 A resulting 2\% variation in the first \cproxy interval
 for both \vtwotwo and \vthreetwo is found. 
 The weighting procedure to account for detector efficiency is also
 scrutinised, by recomputing the efficiency and purity in simulation after increasing  the number of intervals
 in the ($\eta$, \pt) variables by a factor of three. The resulting variations of the flow coefficients 
 are within 2\% for the most central collisions.

To confirm the estimated systematic uncertainty, the stability of
the results against some relevant variables is verified. 
As the acceptance and reconstruction efficiency is highly dependent on the 
longitudinal PV position, the analysis is performed 
for events originating from the first or second half of the SMOG2 cell.
Systematic shifts of the measured values within 2\% are observed, 
consistent with the quoted systematic uncertainties.
A similar test is performed by shifting backward the $|\Delta \eta|=1.0$ excluded region by 0.2 units. 
The \vtwotwo results are found to be stable within 2\%.
To verify that the argon contamination in the neon dataset is correctly
accounted for, the PbNe flow coefficients are recomputed by only
considering a fraction of the total data, acquired long after any other
argon injection. The $v_n$ results in the most affected 0-3\% interval  
change by less than 1\%, indicating that
the assigned systematic uncertainty is a conservative estimate.

\begin{figure}[tb]
    \centering
    \begin{minipage}[t]{1.\linewidth}
        \centering
        \includegraphics[width=\linewidth]{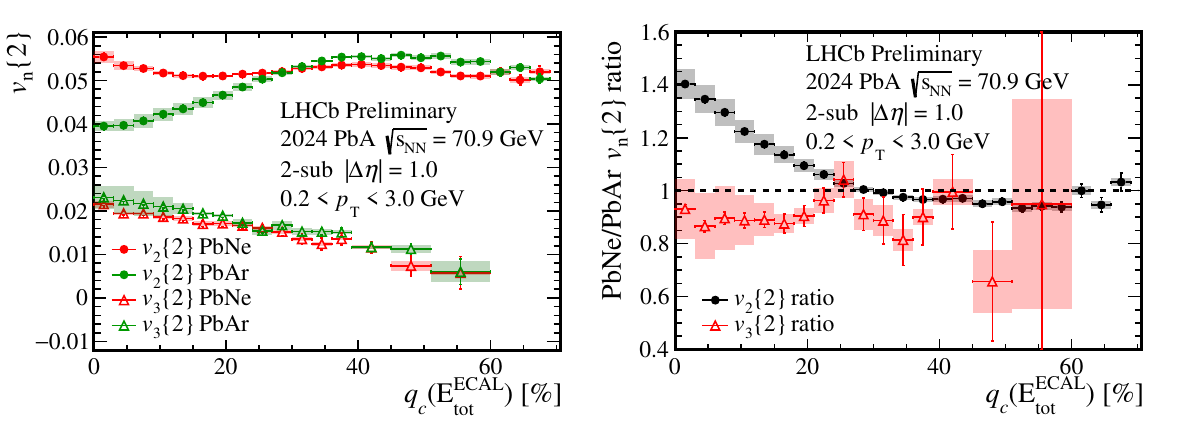}
    \end{minipage}
    \vspace{1em}
    \vspace{1em}
    \begin{minipage}[t]{0.49\linewidth}
        \centering
        \includegraphics[width=\linewidth]{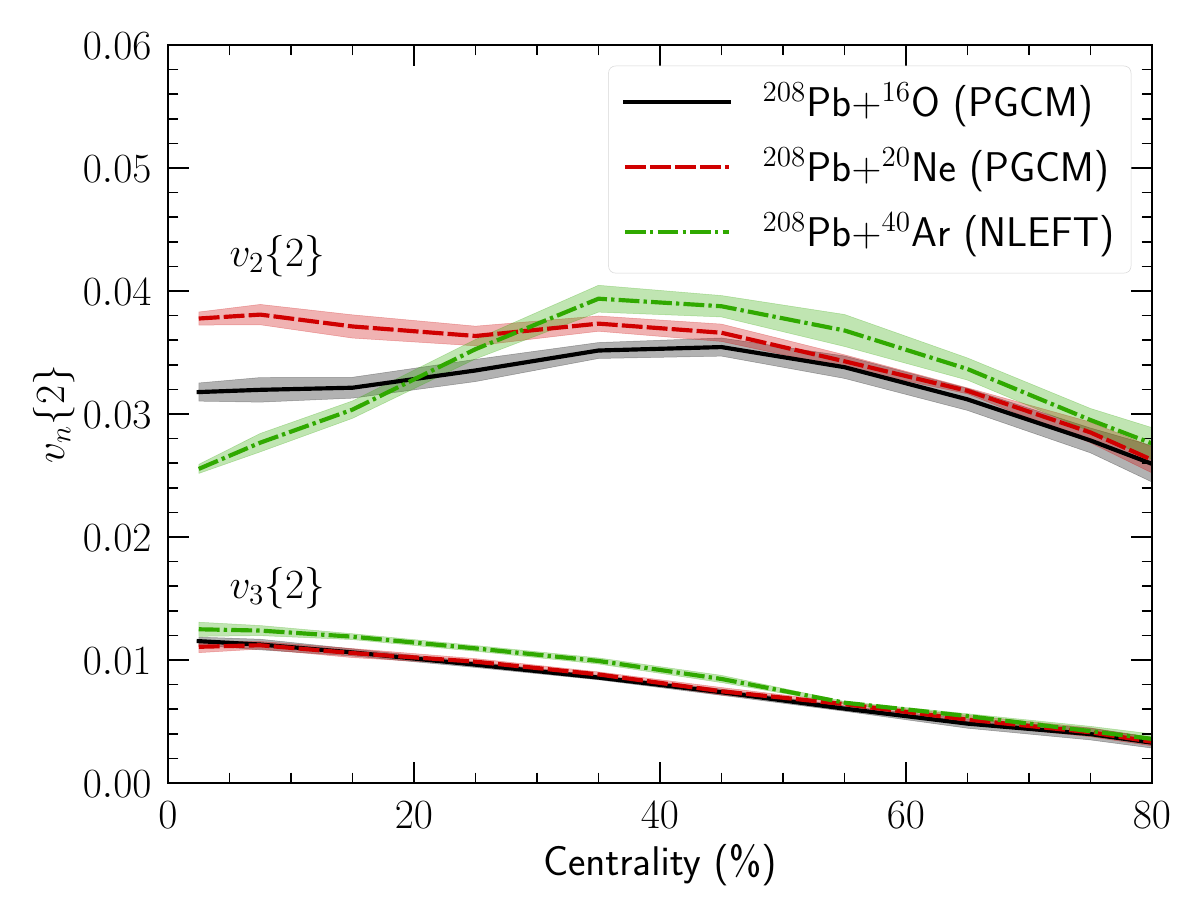}
    \end{minipage}
    \begin{minipage}[t]{0.49\linewidth}
        \centering
        \includegraphics[width=\linewidth]{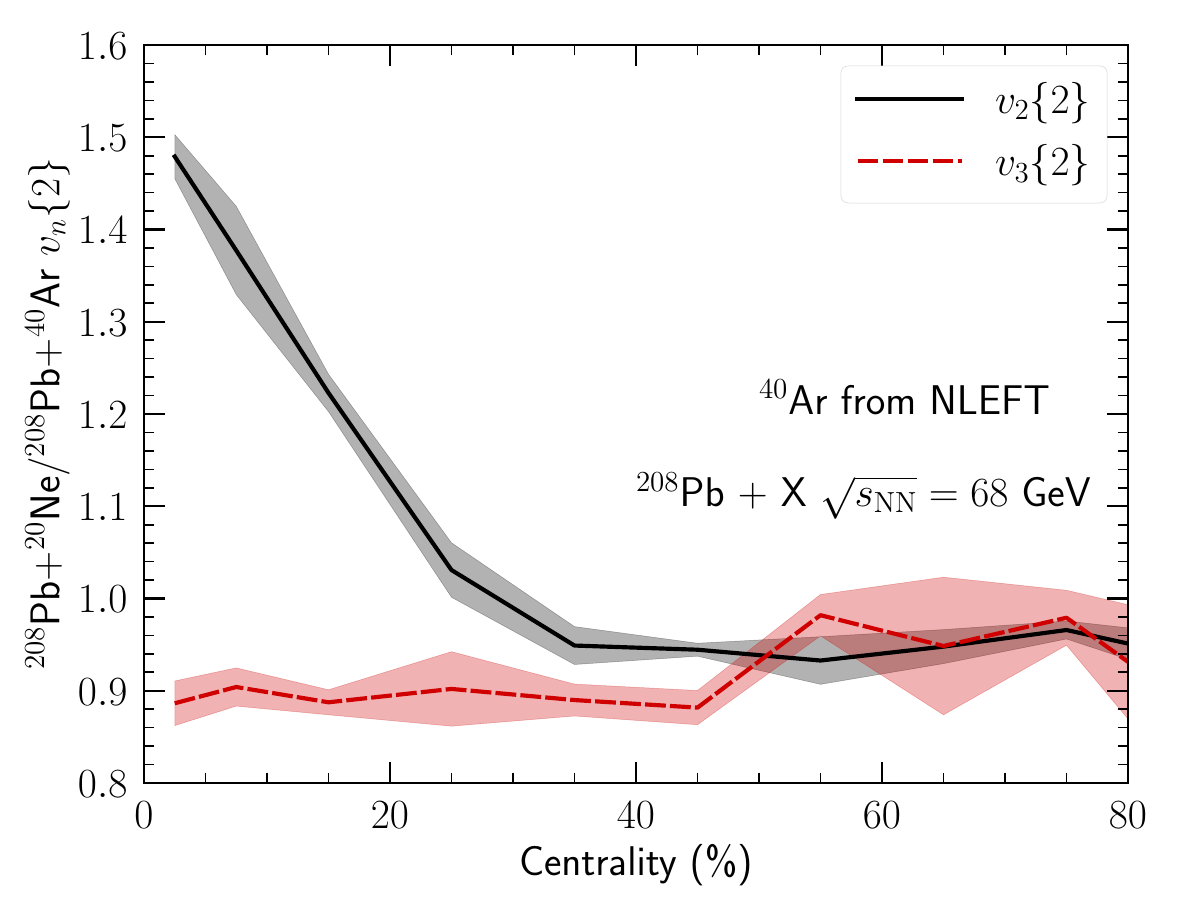}
    \end{minipage}
    \caption{Results for the measurement of the (top left)~\vtwo and \vthree flow 
      coefficients as a function of the centrality proxy \cproxy in
      PbNe and PbAr collisions, along with their (top right)~ratio between PbNe and PbAr collisions. The error bars indicate the statistical
      uncertainty and the shaded boxes the total systematic uncertainty, 
      resulting from the quadratic sum of the single contributions discussed
      in the text. Plots taken from Ref.~\cite{QM25_CHUN_new} with (bottom left)~3+1D hydrodynamic predictions and \textit{ab-initio} nuclear structure inputs, together with (bottom right) their ratio between PbNe and PbAr collisions, are also shown for comparison. }
    \label{fig:vn_comparison}
\end{figure}

Figure~\ref{fig:vn_comparison} presents the preliminary results for the measurement
of the \vtwotwo and \vthreetwo flow coefficients on PbNe and
PbAr fixed-target data at $\sqsnn = 70.9\gev$, as a function of \cproxy. 
The results are obtained with the subevent two-particle cumulant
method in the pseudorapidity ranges $2.4<\eta<3.5$  and $4.5<\eta<5.0$, corresponding to rapidities in the nucleon-nucleon centre-of-mass system of $-1.9<y<-0.8$  and $0.2<y<0.7$, respectively, 
and are integrated over the transverse momentum range
$0.2 < \pt < 3.0 \gev$.

Clearly distinct behaviour between the two collision systems is
observed for the \vtwotwo values at low centrality, with a value
larger by a factor $1.40\pm 0.06$ in PbNe than in PbAr collisions
in the first $0<\cproxy<3\%$  interval.
For the triangular flow coefficient \vthreetwo, both systems
exhibit a decreasing trend with increasing centrality, and PbAr values
are  larger than those of PbNe throughout the explored
centrality range. 

The results are compared  with recent 3+1D hydrodynamic
predictions~\cite{Giacalone:2025prl,QM25_CHUN_new} including \textit{ab-initio} nuclear-structure
inputs, shown in the bottom plots of Fig.~\ref{fig:vn_comparison}. 
These calculations anticipated the observed large difference between 
the \vtwotwo values in PbNe and PbAr collisions, originating from the distinctive shape of the
$^{20}$Ne nucleus. 
It should be stressed that a quantitative comparison with these
predictions is limited by the different assumptions in the
calculations: the cumulant values are predicted for  \mbox{$2.0<\eta<5.0$} without any excluded region in pseudorapidity, and for a lower centre-of-mass
energy of 68\gev.  
A pure $^{20}$Ne target is assumed, while $9\%$ of Ne target
particles in data are $^{22}$Ne nuclei, which are expected to have a more symmetric shape~\cite{PhysRevC.84.034313}. Besides, the nonflow short-range two-particle correlations are not accounted for in the predictions.
It is also worth noting that the accuracy of the centrality proxy
\cproxy in data has still to be evaluated, and that in the predictions two
different models are used for the neon and argon shapes,
PGCM~\cite{PGCM,Frosini:2021sxj,Frosini:2021ddm} and NLEFT~\cite{Lee:2008fa,Lahde:2019npb,NLEFT}, respectively.

Taking these caveats into account, the data provide clear confirmation
of the predicted distinctive signature of the bowling-pin shape of the
 $^{20}$Ne nucleus, reflected in the centrality dependence of
 the elliptic and triangular flow in PbNe collisions.
The ratio $\vtwotwo(\text{PbNe})/\vtwotwo(\text{PbAr})$, where many theoretical
and experimental uncertainties cancel, is found to be in
good agreement with the prediction.

\begin{table}[tb]
    \centering
    \caption{Summary of the relative systematic uncertainties affecting the \vtwotwo and \vthreetwo flow-coefficient measurement in the fixed-target PbNe and PbAr collision samples, given in percent. }
    \small
    \begin{tabular}{lrrrrrrrr}
    \toprule
         \multirow{3}{*}{Systematic source} &  \multicolumn{8}{c}{Systematic uncertainty}\\
         \addlinespace
         & \multicolumn{4}{c}{PbNe} & \multicolumn{4}{c}{PbAr} \\
         \addlinespace
         & \multicolumn{2}{c}{\vtwotwo} & \multicolumn{2}{c}{\vthreetwo} & \multicolumn{2}{c}{\vtwotwo} & \multicolumn{2}{c}{\vthreetwo}\\
         \multicolumn{1}{r}{\cproxy [\%]}  
                           & 0--3 & 36--39  & 0--3  & 36--39 & 0--3  & 36--39  & 0--3  & 36--39 \\
         \midrule
    Ghost tracks rejection &    2.4&   0.9    &  3.1   &   1.2   &  2.5   &   0.4    & 12.5   &   3.3    \\
    \addlinespace
    Clone tracks rejection &    0.9&   0.4   &   1.0  &    1.0  &   1.6   &    0.5   &  1.4   &    0.2    \\
    \addlinespace
    Secondary tracks rejection& $<0.1$&   $<0.1$    &  $<0.1$   &   $<0.1$   &  $<0.1$   &   $<0.1$    & $<0.1$    &   0.4      \\
    \addlinespace
    PbNe argon contamination &  1.2&   $<0.1$    &  2.2   &   $<0.1$   &   --   &   --     &   --   &     --     \\ 
    \addlinespace
    Track weighting procedure & 0.6&   0.4    &  0.4   &   0.7   &  1.6   &   0.2    & 0.9    &   1.5      \\ 
     \midrule                 
     Total                 &    2.9&   1.1    &  4.0   &   1.7   &  3.3   &   0.6    & 12.6   &   3.7    \\
    \bottomrule
    \end{tabular}
    \label{tab:systematics}
\end{table}

 %%%%%%%%%%%%%%%%%%%%%%%%%%%%%%%%%%%%%%%%%%%%%%%%%%%%%%%%%%%%%%%%%%%%%%%%%%%%%%%%%%%%%%%%%%%%%%%%%
\section{Conclusions and outlook}
\label{sec:closing}
In this note, the first experimental evidence of the
effect of the bowling-pin shape of the $^{20}$Ne nucleus 
on the anisotropic flow in PbNe collisions is presented.
Via the multiparticle-correlation-cumulant method, the dependence of \vtwo
and \vthree on centrality via a proxy is measured using the unique LHCb SMOG2
 datasets of fixed-target PbNe and PbAr collisions at $\sqsnn =
70.9\gev$, collected in 2024.   
As the first use of the cumulant method at LHCb, it is
first validated on 2018 PbPb collisions, with which previous \vtwo and \vthree
results obtained with the two-particle correlation method are
reproduced. 

The measured anisotropic flow coefficients \vtwotwo and \vthreetwo 
show distinctly different centrality dependencies in PbNe
and PbAr collisions. In particular, \vtwotwo in PbNe collisions is larger by up to a factor
1.4 with respect to PbAr for the most central collisions. The \vthreetwo
values decrease with centrality in both systems, with PbAr
consistently yielding slightly higher values. Such
nontrivial dependencies are described well by hydrodynamic models incorporating \textit{ab-initio}
nuclear-structure inputs, pointing to a clear signature of the 
peculiar shape of the $^{20}$Ne nucleus and to the validity of the
hydrodynamic description of the hot medium formed in these collisions.

It has to be emphasised that the \lhcb experiment already collected samples of
PbAr and PbNe collisions~\cite{LHCb-PAPER-2022-011} during the LHC Run 2 using the first
version of its gas target SMOG~\cite{FerroLuzzi:2005em, smog}. Those samples were useful for
the early development of this study, but were limited by the
inability of the detector to capture the most central collisions.
The presented results demonstrate that PbAr collisions can be reconstructed over the full centrality range with the \lhcb Upgrade I detector,
due to its  improved granularity.

These results enable precision tests of models for both 
the initial nuclear state and the formation of QGP in
relatively small collision systems. A more quantitative comparison with models can be
made once calculations more closely reflect the experimental conditions, and the
centrality of the collisions in data is estimated more accurately.
Such studies can be extended in the future by exploring 
the dependence of flow coefficients on rapidity and transverse momentum,
their per-event fluctuations
and by considering more observables like four-particle cumulants.

The ongoing fixed-target programme of the \lhcb experiment will allow
more data to be taken with different collisions systems. Beside Ne and
Ar, SMOG2 has already allowed injection of H$_2$, D$_2$, and He gas, while
N$_2$, O$_2$, Kr and Xe gases are under consideration for the future.
The results presented provide initial evidence of the potential for
\lhcb with the SMOG2 target to operate as a unique and powerful tool for nuclear
imaging and for the precise characterisation of the emergence of
collective dynamics in small systems.

% Do not include this in any draft (just for information in the template)
%\input{LHCbInternal/acknowledgements_template}
% Comment this in for paper drafts; do not include this in analysis note, conference and figure reports
%\input{acknowledgements}

\newpage
\addcontentsline{toc}{section}{References}
%\setboolean{inbibliography}{true}
\bibliographystyle{LHCb/LHCb}
\bibliography{main,LHCb/standard,LHCb/LHCb-PAPER,LHCb/LHCb-CONF,LHCb/LHCb-DP,LHCb/LHCb-TDR}

\newpage

\end{document}